\documentclass[11pt]{article}  
\usepackage{amssymb}
\usepackage{latexsym}
\usepackage{a4}

\newtheorem{Theorem}{Theorem}[section]          
\newtheorem{Corollary}[Theorem]{Corollary}      
\newtheorem{Lemma}[Theorem]{Lemma}
\newtheorem{Proposition}[Theorem]{Proposition}  

\newcommand{\Proof}{\noindent{\bf Proof:}\quad}

\newcommand{\qed }
   {\hfill$\hbox{\vrule height1.3ex width1.3ex depth.1ex}\ $
    \par\medskip}

\newcommand{\bref}[1]{(\ref{#1})}

\newcommand{\cM}{{\cal{M}}}          
\newcommand{\cG}{{\cal{G}}}          
\newcommand{\cGp}{\cG^+}             
\newcommand{\cGpl}{\cG^+_\lambda}    

\newcommand{\cK}{{\cal{K}}}          
\newcommand{\cC}{{\mathcal{C}}}          
\newcommand{\cF}{{\mathcal{F}}}          
\newcommand{\cRo}{{\mathcal{R}}_o{}}     
\newcommand{\cPo}{{\mathcal{P}}\kern -0.25em _o{}}     
\newcommand{\cT}{{\mathcal{T}}}          
\newcommand{\cL}{{\mathcal{L}}}          
\newcommand{\cLEH}{{\mathcal{L}}_{EH}}   
\newcommand{\cS}{{\mathcal{S}}}          
\newcommand{\cSt}{\tilde{{\cal{S}}}} 

\newcommand{\bR}{{\mathbb{R}}}       
\newcommand{\bPadm}{{\mathbb{P}}}    
\newcommand{\loc}{\mathrm{loc}}

\newcommand{\madm}{m_{ADM}}

\newcommand{\cHadm}{{\mathcal{H}}_{ADM}} 
\newcommand{\cHrt}{{\mathcal{H}}_{RT}}   
\newcommand{\cH}{{\mathcal{H}}}   

\newcommand{\go}{\mathring{g}}  
\newcommand{\grado}{\mathring{\nabla}}   
\newcommand{\Gammao}{\mathring{\Gamma}}  
\newcommand{\gt}{\tilde{g}}
\newcommand{\gradt}{\tilde{\nabla}}
\newcommand{\trg}{\hbox{tr}_g}       
\newcommand{\tro}{\hbox{tr}_{\go}}   
\newcommand{\rtg}{\sqrt{g}}          
\newcommand{\rto}{\sqrt{\go}}        
\newcommand{\grad}{\nabla}
\newcommand{\del}{\partial}
\newcommand{\Lap}{\Delta}
\newcommand{\Lapo}{\Delta_{\hbox{$\scriptscriptstyle{o}$}}{}}

\newcommand{\Ric}{{Ric}}
\newcommand{\xinf}{\xi_\infty}
\newcommand{\xhinf}{{\hat{\xi}}_\infty}

\newcommand{\nm}[2]{\,\Vert{#1}\Vert_{#2}} 
\newcommand{\len}[1]{\,\vert{#1}\vert}     

\newcommand{\sbullet}{\raise0.3ex\hbox{$\scriptscriptstyle{\bullet}$}}
\newcommand{\half}{{\textstyle{\frac{1}{2}}}}
\newcommand{\tfrac}[2]{{\textstyle{\frac{#1}{#2}}}}   
\renewcommand{\emptyset}{\hbox{\O}}

\begin{document}
\title{Phase Space for the Einstein Equations}
\author{Robert Bartnik \thanks{I would like to thank Vince Moncrief, John Hutchinson, Leon
   Simon and Rick Schoen for valuable discussions.  This research was
   supported in part by the Australian Research Council.}
\footnote{email:  robert.bartnik@canberra.edu.au}
\\
School of Mathematics and Statistics\\
University of Canberra\\
ACT 2601, Australia}
\date{\today}
\maketitle

 \begin{abstract} 
   A Hilbert manifold structure is described for the phase space $\cF$
   of asymptotically flat initial data for the Einstein equations.  The
   space of solutions of the constraint equations forms a Hilbert
   submanifold $\cC \subset \cF$.
   The ADM energy-momentum defines a function which is
   smooth on this submanifold, but which is not defined in general on
   all of $\cF$. The ADM Hamiltonian defines a smooth function on $\cF$
   which generates the Einstein evolution equations only if the
   lapse-shift satisfies rapid decay conditions. However a regularised
   Hamiltonian can be defined on $\cF$ which agrees with the
   Regge-Teitelboim Hamiltonian on $\cC$ and generates the evolution
   for any lapse-shift appropriately asymptotic to a (time) translation
   at infinity. Finally, critical points for the total (ADM) mass,
   considered as a function on the Hilbert manifold of constraint
   solutions, arise precisely at initial data generating stationary
   vacuum spacetimes.
 \end{abstract} 



\maketitle

\section{Introduction}\label{intro:sec}

It has long been known that the Einstein equations can be expressed
as a Hamiltonian field theory, at least formally.  Our aim is to
justify these calculations by providing Hilbert space structures in
which important quantities such as the constraint map and the total
energy-momentum, become smooth functions.
We work with a phase space $\cF$ consisting of pairs $(g,\pi)$ of
$H^2\times H^1$ local regularity with decay appropriate for
asymptotically flat spacetimes.  Our main results imply in particular:
\begin{itemize}
\item the constraint set $\cC$ is a Hilbert submanifold of $\cF$
  (Theorem \ref{constraintmfld:thm});
\item the ADM energy-momentum is a $C^\infty$ function on $\cC$
  (Theorem \ref{Padm-smooth:thm});
\item a regularization $\cH(g,\pi;\xi)$ of the Regge--Teitelboim (RT)
  Hamiltonian is $C^\infty$ on $\cF$ and generates the correct
  equations of motion (Theorem \ref{HRT:thm}); and 
\item constrained critical points of the regularized Hamiltonian $\cH$
  on $\cC$ correspond to Killing initial data (Theorem
  \ref{Ecrit:thm}).
\end{itemize}

In \S\ref{constraint:sec} we show that the set of asymptotically flat
solutions to the constraint equations $\Phi(g,\pi)=0$ is a smooth
Hilbert submanifold of the phase space $\cF$.  This is the property of
\emph{linearization stability} \cite{FischerMarsden79}, so-called
because it implies that any solution of the linearized Einstein
equations corresponds to a curve of solutions of the nonlinear
equations, provided a suitable local existence result is available for
the regularity class in question.

However, the best local existence and uniqueness results for the
vacuum Einstein evolution at present require slightly more:
$(g,\pi)\in H^s\times H^{s-1}$ with $s>2$
\cite{BahouriChemin99,Tataru02,KlainermanRodnianski02}.  Interestingly
it has been conjectured that this can be improved to $s=2$, the case
examined here, and possibly even to $s>3/2$, but the calculations here
rely heavily on $s=2$ and it is not clear whether they can be extended
further.

The ADM total mass and energy momentum \cite{ADM61} are defined by
limits at infinity of coordinate-dependent integrals.  The consistency
of these definitions, and their independence of the coordinate
framing, is established in \S\ref{ADM:sec}; this extends previous
results \cite{Bartnik86,Chrusciel86a,OMurchadha86}.  Furthermore, the
the ADM energy-momentum is a smooth function on the constraint
manifold $\cC$; however, it is not finite in general on $\cF$.

The Einstein evolution equations may be written in Hamiltonian form
\cite{ADM62}, with the lapse-shift $\xi$ freely specifiable.  In
\S\ref{hamiltonian:sec} we show that the ADM Hamiltonian is also
smooth on $\cF$, provided $\xi$ decays suitably, and its derivative on
$\cF$ generates the evolution equations.  To extend this result to
$\xi$ asymptotic to a time translation at infinity we modify the RT
Hamiltonian \cite{ReggeTeitelboim74} to construct a regularized
Hamiltonian which is smooth on all $\cF$ and agrees with the ADM
energy-momentum on $\cC\subset \cF$.  

It is appealing to conjecture that, although the Hamiltonian flow
vector field is only densely defined on $\cF$, it might still be
possible to construct integral curves directly, perhaps by a judicious
choice of lapse-shift $\xi$ to smooth the tangent vectors.  This would
amount to a direct proof of local existence for $s=2$ and is unlikely
to succeed, because it does not take into account the characteristic
structure of the Einstein equations, which plays an important role in
other approaches to the low-regularity local existence problem
\cite{KlainermanRodnianski02,Tataru02}.

The lapse-shift $\xi$ in the regularized Hamiltonian may be regarded
as a Lagrange multiplier for constrained variations, and in
\S\ref{sec:critical} we use this to establish rigorously an identity
of Brill-Deser-Fadeev \cite{BrillDeserFadeev68}, which equates
constrained critical points of the ADM energy with Killing initial
data, i.e. $D\Phi(g,\pi)^*\xi=0$.

Critical points of energy arise naturally from the mass-minimizing
definition of \emph{quasi-local mass} \cite{Bartnik89,Bartnik02},
which motivates the conjecture that mass-minimizing extensions of a
given region $(\Omega,g,\pi)$ are stationary.  The static case has
been established in \cite{Corvino00} by a different method, but a
direct variational proof, based on extending the results of
\S\ref{sec:critical} to data sets with boundary, would be more
natural.  This question will be addressed in future work.

\section{Notation and Formulae}\label{notation:sec}
%
%
In this section we introduce the basic framework and notations used in the
paper, and recall some well-known formulae concerning the constraint
equations.

Let $\cM$ be a connected, oriented and non-compact 3-dimensional
manifold, and suppose there is a compact subset $\cM_0\subset\cM$ such
that there is a diffeomorphism $\phi:\cM \backslash \cM_0 \to E_1$,
where $E_R\subset \bR^3$ is an exterior region, $E_R = \{x\in\bR^3 :
\len{x} > R\}$.  We also use the notation $B_R$ for the open ball of
radius $R$ centred at $0\in\bR^3$, $A_R = B_{2R}\backslash
\overline{B_R}$ for the annulus and $S_R=\partial B_R$ for the sphere
of radius $R$.  Although we assume $\partial M=\emptyset$ for
simplicity, most of the earlier results are valid also when
$\partial\cM$ is non-empty and consists of a finite collection of
disjoint compact 2-surfaces.
 Let $\go$ be a fixed
Riemannian metric on $\cM$ satisfying $\go = \phi^*(\delta)$ in $\cM
\backslash \cM_0$, where $\delta$ is the natural flat metric on
$\bR^3$.  In the terminology of \cite{Bartnik86}, $\phi$ is a {\em
  structure of infinity\/} on $\cM$.  Let $r\in C^\infty(\cM)$ be some
function satisfying $r(x) \ge 1\ \forall x\in\cM$ and $r(x) = |x|\ 
\forall x\in\cM\backslash \cM_0$.  Using $r$ and $\go$ we define the
weighted Lebesgue and Sobolev spaces \cite{Bartnik86} $L^p_\delta$,
$W^{k,p}_\delta$, $1\le p\le\infty$, $\delta\in\bR$, as the
completions of $C_c^\infty(\cM)$ under the norms
\begin{eqnarray*}
   \nm{u}{p,\delta}\   &=& \left(\int_\cM |u|^p r^{-\delta p -3} 
                              dv_o\right)^{1/p}, \\ 
   \nm{u}{k,p,\delta} &=& \sum_{j=0}^k \nm{\grado^j u}{p,\delta-j},
\end{eqnarray*}
if $p<\infty$, and the appropriate supremum norm if $p=\infty$.
Here $dv_o, \grado$ are respectively the volume measure and connection 
determined by  the metric $\go$.
The weighted Sobolev space of sections of a bundle $E$ over $\cM$ is
defined similarly and denoted $W^{k,p}_\delta(E)$.
We distinguish especially the spaces
\[
\begin{array}{rclrcl}
   \cG   &=& W^{2,2}_{-1/2}(\cS),    &\qquad
   \cK   &=& W^{1,2}_{-3/2}(\cSt),   \\ 
   \cL   &=& L^2_{-1/2}(\cT),        &
   \cL^* &=& L^2_{-5/2}(\cT^*\otimes\Lambda^3),
\end{array}
\]
where $\cS = S^2T^*\cM$ is the bundle of symmetric bilinear forms on $\cM$,
$\cSt=S^2T\cM\otimes\Lambda^3T^*\cM$ is the bundle of symmetric
tensor-valued 3-forms (densities) on $\cM$ and $\cT$ is the bundle of
spacetime tangent vectors.  Thus, for example, $\cL$ is a class of
spacetime tangent vector fields on $\cM$, and $\cL$ and $\cL^*$ are dual
spaces with respect to the natural integration pairing.  The following
Hilbert manifolds modelled on $\cG$ are natural domains for asymptotically
flat metrics:
\begin{eqnarray*}
   \cGp  &=& \{ g : g-\go \in \cG, g > 0\} ,              \\
   \cGpl &=& \{ g\in\cGp,  \lambda\go <  g < \lambda^{-1} \go \},
          \quad  0 < \lambda < 1. 
\end{eqnarray*}
We note that by virtue of the Sobolev inequality and the Morrey lemma 
\cite{Bartnik86}, 
tensors in $\cG$ are H\"older continuous (with H\"older exponent $1/2$) 
and thus the matrix inequality conditions on  $g$ in the definitions of $\cGp, 
\cGpl$ are satisfied in the pointwise sense.
The Hilbert manifold we shall consider as the phase space for the 
Einstein equations is then
\begin{equation}
        \cF = \cGp\times\cK .
        \label{F:defn}
\end{equation}
Theorem \ref{ADM-uniqueness:thm} shows that $\cF$ is independent of the 
choice of structure of infinity $\phi$.

If we suppose that $\cM$ is a spacelike submanifold of a 4-dimensional 
Lorentzian manifold (spacetime), then 
the second fundamental form or extrinsic curvature tensor $K$ is the 
bilinear form defined by 
\begin{equation}
    K(u,v) = g^{(4)}(u,\grad^{(4)}_v n),
\end{equation}
where $g^{(4)}$, $\grad^{(4)}$ are the spacetime metric and connection,
$u,v$ are tangent vectors to $\cM$ and $n\in\cT$ is the future unit 
normal to $\cM$.
It is often convenient to use the conjugate momentum 
$\pi$ as a reparameterisation of $K$ --- we adopt the definition
\begin{equation}\label{pi:def}
   \pi^{ij} = (K^{ij} - \trg{K}\,g^{ij} ) \rtg,
\end{equation}
where $\rtg=\sqrt{\det{g}}/\sqrt{\det{\go}}$ 
denotes the volume form of the induced metric $g$, so $\pi$ 
is a section of the bundle $\cSt = S^2T\cM\otimes\Lambda^3T^*\cM$.
Either $(g,K)$ or $(g,\pi)$ can be used as coordinates on $\cF$,
and we will move freely between these two parameterisations in 
the following formul\ae.

For sufficiently smooth metric $g$ and second fundamental form $K$ 
(or $\pi$), 
the constraint functions $\Phi = (\Phi_0,\Phi_i) = \Phi(g,\pi)$ are 
defined by
\begin{eqnarray}
\label{Phi0:def}
   \Phi_0(g,\pi) &=& \left(R(g) - \len{K}^2 + (\trg{K})^2\right)\rtg,   \nonumber \\
                 &=& R(g)\rtg - (\len{\pi}^2 - \half(\trg\pi)^2)/\rtg  
  \\[3pt]
\label{Phii:def}
   \Phi_i(g,\pi) &=& 2\, \left(\grad^j K_{ij} - \grad_i (\trg{K})\right)\rtg  
              \nonumber \\
                 &=& 2\, g_{ij}\grad_k \pi^{jk},
\end{eqnarray}
where $R(g)$, $\grad$, $\trg$ are respectively the Ricci scalar, covariant 
derivative and trace of the metric $g$, and 
$\len{K}^2 = g^{ik}g^{jl}K_{ij}K_{kl}$.
Notice that $\Phi$ takes values in $\cT^*\otimes\Lambda^3T^*\cM$, 
the bundle of density-valued spacetime cotangent vectors on $\cM$.
If the Einstein equations are satisfied then the normalisation chosen
ensures that $\Phi$ and the stress-energy tensor are related by
$\Phi_\alpha = 16\pi\kappa T_{n\alpha}\rtg$, 
where $n=e_0$ is the future unit normal to $\cM$,
$\kappa$ is Newton's gravitational constant,
and $T_{n\alpha}\rtg$ is the local energy-momentum density 
$4$-covector as seen by an observer with world vector $n$.
Consequently our sign conventions vary slightly from those used in 
\cite{Moncrief75,FischerMarsden79}.

The functional derivative $D\Phi$ is given 
formally by
\begin{eqnarray}
   \lefteqn{D\Phi_0(g,\pi)(h,p)\ \ = } \qquad       \nonumber\\
    && \left(\delta_g\delta_g h - \Lap_g\trg h\right)\rtg
          - h_{ij}\left( \Ric^{ij}- \half R(g)g^{ij}\right)\rtg  \nonumber \\
    &&{}+ h_{ij}\left( \trg\pi\,\pi^{ij} - 2\pi^i_k\pi^{kj}
                 +\half\len{\pi}^2 g^{ij} 
                 -\tfrac{1}{4}(\trg\pi)^2g^{ij}
                  \right)/\rtg                                  \nonumber \\   
         &&{}+ p^{ij}\left(\trg\pi g_{ij}-2\pi_{ij}\right)/\rtg,
\label{DPhi0:def} \\[3pt]
\lefteqn{D\Phi_i(g,\pi)(h,p)\ =\ 
             \pi^{jk}\left(2\grad_j h_{ik} - \grad_i h_{jk}\right)
              + 2 h_{ij}\grad_k \pi^{jk}  + 2g_{ik}\grad_j p^{jk},
            }\qquad 
\label{DPhii:def}
\end{eqnarray}
where $\delta_g\delta_g h = \grad^{i}\grad^jh_{ij}$.  Multiplying by
$(N,X^i)$ and integrating by parts and ignoring boundary terms gives
formul\ae\ for the formal $L^2(dv_o)$-adjoint operator $D\Phi^*$,
\begin{eqnarray}
\lefteqn{(h,p)\!\cdot\! D\Phi_0(g,\pi)^*N \ \ =} \qquad \nonumber \\
  &&  h_{ij}\,\left(\grad^{i}\grad^jN - \Lap_g Ng^{ij}\right)\rtg  
     - N h_{ij}\,\left( \Ric^{ij}- \half  R(g)g^{ij}\right)\rtg  \nonumber \\
  &&{}+ N h_{ij}\,\left( \trg\pi\,\pi^{ij} - 2\pi^i_k\pi^{kj}
                        +\half\len{\pi}^2 g^{ij} 
                        -\tfrac{1}{4}(\trg\pi)^2g^{ij}
                \right)/\rtg                                  \nonumber \\   
  &&{}+ N p^{ij}\,\left(\trg\pi g_{ij}-2\pi_{ij}\right)/\rtg,
\label{DPhi0*} \\[3pt]
\lefteqn{(h,p)\!\cdot\!D\Phi_i(g,\pi)^*X^i\ \ =} \qquad \nonumber\\
  && h_{ij}\,\left( X^k\grad_k\pi^{ij} + \grad_kX^k \pi^{ij}
                     -2\grad_kX^{(i}\pi^{j)k}  \right)            
      - 2\, p^{ij}\grad_{(i}X_{j)}.
\label{DPhii*}
\end{eqnarray}
These calculations are carefully described in \cite{FischerMarsden79}.
Adopting some natural shorthand notations, the adjoint operator 
can be rewritten
\begin{eqnarray}
\lefteqn{(h,p)\!\cdot\!D\Phi(g,\pi)^*(N,X)\ \  = } \qquad \nonumber \\
  && h\,\sbullet\biggl\{
   \left(\grad^2N-\Lap_gN g    
   -N\left({\Ric}-\half R(g)g\right)\right)^{\scriptscriptstyle\#}\!\rtg  
\nonumber 
\\
  &&\quad   - N\left(K\pi+\pi K - \half\pi\,\sbullet K\,g
             \right)^{\scriptscriptstyle\#}  + \cL_X\pi \biggr\} 
   {} - p\,\sbullet\left(2KN+\cL_Xg\right)
\label{DPhi*}
\end{eqnarray}
where ${}^{\scriptscriptstyle\#}$ signifies the indexed-raised tensor, 
$\cL_X$ is the Lie derivative in the direction $X$, 
$(K\pi)^{ij} = K^i_k\pi^{kj}$
and $\sbullet$ is the natural contraction between 2-tensors, 
eg.~$\pi\,\sbullet K = \pi^{ij}K_{ij}$.  Defining 
\begin{equation}
   S^{ij} = g^{-1} (\trg\pi \pi^{ij} -2\pi^i_k\pi^{jk} 
      +\half|\pi|^2g^{ij}-\tfrac{1}{4}(\trg\pi)^2)g^{ij}),
\end{equation}
and $E^{ij}=\Ric^{ij}-\half R(g)g^{ij}$, we may express $D\Phi$  in matrix
form as 
\begin{equation}
\label{DPhi}
   D\Phi(g,\pi)(h,p) = \left[ \begin{array}{cc}
    \sqrt{g}(\delta_g\delta_g-\Lap_g\trg +S-E) & - 2K \\
    \hat{\pi}\nabla + 2\delta_g\pi  & 2\delta_g \end{array}\right]
   \left[ \begin{array}{c} h \\ p \end{array} \right],
\end{equation}
where 
\[
\hat{\pi}\nabla h = \hat{\pi}^{jkl}_i \nabla_jh_{kl}
  = (\pi^{jk}\delta_i^l + \pi^{jl}\delta_i^k
-\pi^{kl}\delta_i^j)\nabla_jh_{kl}.
\]
Similarly the adjoint may be written as
\begin{equation}
\label{DPhi*1}
   D\Phi(g,\pi)^*(N,X) =
\left[ \begin{array}{cc} 
       \sqrt{g}(\nabla^2- g \Lap_g + S-E) & \nabla \pi-\hat{\pi}\nabla   \\
       -2K & -\epsilon_g \end{array} \right]
  \left[ \begin{array}{c} N \\ X 
         \end{array}\right],
\end{equation}
where 
\[
(\nabla \pi-\hat{\pi}\nabla )X = \cL_X\pi = \nabla_X
\pi^{ij}-\hat{\pi}_l^{kij}\nabla_kX^l
\]
and $\epsilon_g(X)=\cL_Xg=2\nabla_{(i}X_{j)}$ is the strain operator.
Let $D\Phi(g,\pi)^*_a(N,X)$, $a=1,2$ denote the two components of $D\Phi^*$
in \bref{DPhi*1}.

We will also use the notation $\xi = (\xi^\alpha) = (N,X^i)$, where $\xi$
has a natural interpretation as the lapse-shift of the spatial slicing of
the evolved spacetime.  If $g$ and $(N,X^i)$ depend on an evolution
parameter $t$ and $N>0$, then the Lorentzian metric
\begin{equation}
        ds^2 = -N^2 dt^2 + g_{ij}(dx^i+X^idt)(dx^j+X^jdt)
\end{equation}
describes a spacetime satisfying some form of the Einstein equations,
and $\xi=N n + X^i\partial_i$ coincides with the time evolution vector 
$\partial_t$.

Greek letters $\alpha,\beta,\dots$ will be used for spacetime indices, with
range $0,\dots,3$, and Latin letters $i,j,\dots$, will indicate spatial
indices (on $\cM$), with range $1,2,3$.  Index-free and indexed expressions
will be intermixed as convenient.  The letters $c,C$ will be used to
indicate constants which may vary from line to line, with $c$ generally
denoting a constant depending only on the background metric $\go$ and the
ellipticity $\lambda$, and $C$ denoting a constant whose dependence on
significant parameters will be explicitly indicated.

\section{The constraint manifold}\label{constraint:sec}
%
%

In this section we show that the constraint map
\begin{equation}
   \Phi : \cF \to \cL^*
   \label{constraint-map}
\end{equation}
is a smooth map between Hilbert manifolds, and that the level sets
$\cC(\epsilon,S) = \Phi^{-1}(\epsilon,S)$ are Hilbert submanifolds.
In particular, the space $\cC = \Phi^{-1}(0)$ of asymptotically flat
vacuum initial data is a Hilbert manifold.  The proof is based on the
implicit function theorem method used in previous studies
\cite{FischerMarsden79,Moncrief75,ArmsMarsdenMoncrief82} of the
constraint set over a compact manifold.  In fact, the main result of
this section may be considered as the logical extension of those
results to the case of asymptotically flat manifolds.  We note in
particular that the quadratic Taub constraints on the linearised
solutions which arise in the case where the underlying spacetime
admits a symmetry, do not occur in the asymptotically flat case --- as
was observed by Moncrief \cite{Moncrief76} --- and consequently, the
cone-like singularities which occur in the space of solutions of the
constraints over a compact manifold (at data sets generating vacuum
spacetimes admitting a Killing vector), are absent in the
asymptotically flat constraint manifold.  The space of asymptotically
flat (vacuum) constraint data is a smooth Hilbert manifold, at all
points.

However the result shown here, that the space of solutions of the
constraint equations forms a Hilbert manifold, does not prove that the
Einstein equations with asymptotically flat data are linearization
stable, in the sense of \cite{FischerMarsden79,Moncrief75}, because
the regularity condition $(g,\pi)\in\cF$ is too weak to be able to
apply known local existence and uniqueness theorems for the Einstein
equations.  It is interesting, therefore, that it has been conjectured
that the minimal regularity conditions for the well-posedness of the
Einstein equations exactly correspond to $(g,\pi)\in\cF$.  If this
conjecture is correct, then linearization stability will hold under
the conditions considered here as well.

Alternatively, linearization stability may be obtained by requiring
higher differentiability in the spaces $\cGp, \cK$ and $\cL^*$, and
then observing that the results about the boundedness and smoothness
of $\Phi$ and the triviality of the kernel of $D\Phi^*$ remain valid
--- the result is a phase space of initial data with sufficient
regularity for known existence and uniqueness theorems to apply.  The
details of this extension are left to the interested reader.

\begin{Proposition}\label{Phi-L2bnd:thm}
Suppose $g\in\cGpl$ for some $\lambda > 0$ and $\pi\in\cK$. 
Then there is a constant $c=c(\lambda)$ such that
\begin{eqnarray}  
   \nm{\Phi_0(g,\pi)}{2,-5/2} 
       &\le& c\,\bigl( 1+\nm{g-\go}{2,2,-1/2}^2 + \nm{\pi}{1,2,-3/2}^2 \bigr),  
\label{Phi0-L2:bnd}  \\
   \nm{\Phi_i(g,\pi)}{2,-5/2} 
       &\le& c\,\bigl(\nm{\grado \pi}{2,-5/2} 
           + \nm{\grado g}{1,2,-3/2} \nm{\pi}{1,2,-3/2} \bigr).
\label{Phii-L2:bnd}  
\end{eqnarray}
\end{Proposition}

\Proof
Since $g\in\cGpl$, $g$ is H\"older-continuous with H\"older exponent 
$1/2$, and we have the global pointwise bounds
\begin{equation}
        \lambda\go{}_{ij}(x) v^iv^j < g_{ij}(x) v^iv^j 
                                    < \lambda^{-1}\go{}_{ij}(x) v^iv^j
          \quad  \forall\,x\in \cM,\  v\in\bR^3.
        \label{glambda:bnd}
\end{equation}
For later use we note the following consequence of the weighted H\"older 
and Sobolev inequalities \cite{Bartnik86}, valid for any function or tensor field $u$,
\begin{eqnarray}
        \nm{u^2}{2,-5/2} & = & \nm{u}{4,-5/4}^2 \ \le\ c \nm{u}{4,-3/2}^2  \nonumber \\
         & \le & c \nm{u}{6,-3/2}^{3/2} \nm{u}{2,-3/2}^{1/2}  \nonumber \\
         & \le & c \nm{u}{1,2,-3/2}^2.
           \label{u2:bnd}
\end{eqnarray}

The $g,\go$ connections are related by the  difference tensor 
$A_{ij}^k = \Gamma_{ij}^k - \Gammao_{ij}^k$, which may be defined 
invariantly by 
\begin{equation}
A_{ij}^k = \half g^{kl} \bigl(
   \grado_ig_{jl} + \grado_jg_{il} - \grado_lg_{ij}\bigr).
   \label{Aijk:def}
\end{equation}
The scalar curvature can be expressed in terms of $\grado$ and $A_{ij}^k$ by
\begin{eqnarray}
R(g) & = & g^{jk}\Ric(\go)_{jk}
          + g^{jk}\bigl(\grado_iA_{jk}^i - \grado_jA_{ik}^i
                        +A_{jk}^l A_{il}^i - A_{jl}^iA_{ki}^l\bigr)
\nonumber \\
 & = & g^{ik}g^{jl}\bigl(\grado^2_{ij}g_{kl} - \grado^2_{ik}g_{jl}\bigr)
       +Q(g^{-1},\grado g) + g^{jk}\Ric(\go)_{jk},
\label{Rg-go:eqn} 
\end{eqnarray}
where $Q(g^{-1},\grado g)$ denotes a sum of terms quadratic in 
$g^{-1},\grado g$.  
Using \bref{glambda:bnd},\bref{u2:bnd},\bref{Rg-go:eqn} 
we may estimate
\begin{eqnarray*}
\nm{R(g)}{2,-5/2}^2 
 & \le & c \int_\cM
    \bigl(\len{\grado^2g}^2 + \len{\grado g}^4 + \len{\Ric(\go)}^2\bigr)
      r^2\, dv_o  \\
 & \le & c\, \bigl(1+\nm{\grado^2g}{2,-5/2}^2  
                     +\nm{\grado g}{4,-5/4}^4 \bigr)  \\
 & \le & c\, \bigl(1+\nm{\grado g}{1,2,-3/2}^4  \bigr),
\end{eqnarray*}
and since
$$
\nm{\len{\pi}^2}{2,-5/2} \le c \nm{\pi}{1,2,-3/2}^2,
$$
the estimate \bref{Phi0-L2:bnd} follows and $\Phi_0(g,\pi) \in L^2_{-5/2}$.

The proof of the corresponding estimates for the momentum constraint is 
similar but somewhat simpler.
Since 
\begin{equation}
        \grad_j\pi^{ij} = \grado_j\pi^{ij} + A_{jk}^i\pi^{jk},
        \label{divpi-go:eqn}
\end{equation}
we have
\begin{equation}
        \Phi_i(g,\pi) = 2g_{ij}\left(\grado_k\pi^{jk} + A^j_{kl}\pi^{kl}\right),
        \label{Phii-go:def}
\end{equation}
and H\"older's inequality, \bref{glambda:bnd} and \bref{u2:bnd} give
$$
        \nm{\Phi_i(g,\pi)}{2,-5/2}^2 \le c
        \left(\nm{\grado\pi}{2,-5/2}^2 + 
              \nm{\grado g}{1,2,-3/2}^2 \nm{\pi}{1,2,-3/2}^2\right).
$$
\qed

Thus $\Phi$ is a quadratically bounded map between the Hilbert manifolds
$\cF=\cGp\times\cK$ and $\cL^* = L^2_{-5/2}(\cT)$; together with the
polynomial structure of the constraint functionals, this enables us to show
that $\Phi$ is smooth, in the sense of infinitely many Frech\'et derivatives.

\begin{Corollary}
$   \Phi : \cF \to \cL^*  $
is a smooth map of Hilbert manifolds.
\end{Corollary}

\Proof
Proposition  \ref{Phi-L2bnd:thm} shows that  
$ \nm{\Phi(g,\pi)}{\cL^*} \le c(1+ \nm{g-\go}{\cG}^2+\nm{\pi}{\cK}^2)$, 
so $\Phi$ is locally bounded on $\cF$.  
To show $\Phi$ is smooth, we note from the representations
\bref{Aijk:def}, \bref{Rg-go:eqn}, \bref{Phii-go:def}
that $\Phi$ 
can be expressed as the composition
\[
\Phi(g,\pi) = F(g,g^{-1},\sqrt{g},1/\sqrt{g},
                \grado g,\grado^2g,\pi,\grado \pi),
\]
where $F=F(a_1,\dots,a_8)$ is a polynomial function which is quadratic in
the parameters $a_5$ and $a_7$ and linear in the remaining parameters.  The
map $g \mapsto (g,g^{-1},\sqrt{g},1/\sqrt{g})$ is analytic on the space of
positive definite matrices, and the maps $g \mapsto \grado g$, $g \mapsto
\grado^2 g$ and $\pi\mapsto \grado\pi$ are bounded linear, hence smooth,
from the Hilbert manifolds $\cGp$ and $\cK$ to $\cL^*$.  Results of Zorn
and Hille \cite[\S3, \S26]{HillePhillips57} on locally bounded polynomial
functionals show $\Phi$ has continuous Frech\'et derivatives of all orders.
\qed


The constraint set $\cC = \Phi^{-1}(0) \subset\cF$ is of particular 
interest, since it gives the class of initial data for the vacuum 
Einstein equations.  To show that $\cC$ is a Hilbert manifold using the 
implicit function method, we study the kernel of the adjoint operator 
$D\Phi(g,\pi)^*$. 

The first step establishes coercivity of $D\Phi(g,\pi)^*$.

\begin{Proposition} \label{DPhi*:coercive}
$D\Phi^*$ satisfies the ellipticity estimate, for all $\xi\in
W^{2,2}_{-1/2}$,
\begin{equation} \label{xi22}
   \nm{\xi}{2,2,-1/2} \le c(\nm{D\Phi(g,\pi)^*_1(\xi)}{2,-5/2}
  +  \nm{D\Phi(g,\pi)^*_2(\xi)}{1,2,-3/2}) + C\nm{\xi}{1,2,0}
\end{equation}
where $C$ depends on $\go$, $\lambda$ and $\nm{(g,\pi)}{\cF}$.
\end{Proposition}

\Proof
Rearranging the first component of \bref{DPhi0*} gives
\begin{eqnarray}
   \nabla^2N &=& Q-\half\trg Q\,g ,
\label{D2N}
\\
Q &=& D\Phi(g,\pi)^*_1(\xi)/\sqrt{g} + (E-S)N -\cL_X\pi/\sqrt{g},
\nonumber
\end{eqnarray}
and thus $\len{\nabla^2N}^2 \le\tfrac{5}{4}\len{Q}^2$.  This leads to the estimate 
\begin{eqnarray}
   \nm{\grado^2N}{2,-5/2} &\le&
   c\,\biggl(\nm{D\Phi_0(g,\pi)^*_1(\xi)}{2,-5/2} 
\nonumber
\\
   &&  {}+\nm{N}{\infty,0}\,(\nm{E}{2,-5/2} +\nm{S}{2,-5/2} )
     +\nm{A\grado N}{2,-5/2}
\nonumber
\\
   && {}+\nm{X}{\infty,0}\nm{\grado \pi}{2,-5/2} +\nm{\grado
   X}{3,-1}\nm{\pi}{6,-3/2}\biggr).
\label{D2N-a:est}
\end{eqnarray}
Using a combination of the weighted Sobolev and H\"older inequalities we
can establish estimates which control the various right hand terms in
\bref{D2N-a:est}.  For example,
\begin{eqnarray}
   \nm{u}{\infty,0} &\le& c \nm{u}{1,4,0}
\nonumber \\ 
  &\le& c\nm{u}{1,2,0}^\lambda \nm{u}{1,6,0}^{1-\lambda},\qquad
\lambda=\tfrac{1}{4} ,
\nonumber \\  
  &\le& c\nm{u}{1,2,0}^\lambda \nm{u}{2,2,0}^{1-\lambda}
\nonumber \\ 
  &\le& \epsilon \nm{\grado^2u}{2,-2}+c\epsilon^{-3} \nm{u}{1,2,0},
\label{uinf:est}
\end{eqnarray}
for any $\epsilon>0$.  Similarly we find, for any $\delta\in\bR$,
\[
  \nm{u}{3,\delta}\le\epsilon\nm{\grado u}{2,\delta-1}
   +c\epsilon^{-1}\nm{u}{2,\delta}.
\]
Consequently there is a constant $C$, depending only on $\lambda$, $\go$,
$\epsilon$ and $\nm{(g,\pi)}{\cF}$, such that
\begin{equation}
\label{D2N:est}
   \nm{\grado^2N}{2,-5/2} \le
   c \nm{D\Phi_0(g,\pi)^*_1(\xi)}{2,-5/2} +\epsilon\nm{\grado^2\xi}{2,-2}
   +C\nm{\xi}{1,2,0} .
\end{equation}
From the identity (using the metric $g\in\cGp$)
\begin{equation}
        X_{i|jk} = - R_{ijkl}X^l + X_{(i|j)k} + X_{(i|k)j} - X_{(j|k)i},
        \label{Xijk:eqn}
\end{equation}
which is valid for any sufficiently smooth $X_i$, we may write
\begin{equation}
        X_{i|jk} = - R_{ijkl}X^l 
          -\half\left(H_{ij|k}+H_{ik|j}-H_{jk|i}\right)
          - (NK_{ij})_{|k} - (NK_{ik})_{|j} + (NK_{jk})_{|i},
        \label{Xijk:idn}
\end{equation}
where 
\[
H_{ij} = H_{ij}(X) = -2(N K_{ij}+ X_{(i|j)})=D\Phi(g,\pi)^*_2(\xi)
\]
and $(N,X^i)$ are assumed sufficiently smooth.
The various terms of \bref{Xijk:idn} can be controlled using the Sobolev, 
H\"older and interpolation inequalities in a similar fashion, leading to 
the estimate
\begin{equation}
  \nm{\grado^2 X}{2} \le  c \nm{D\Phi_0(g,\pi)^*_2(\xi)}{1,2,-3/2}
   +\epsilon \nm{\grado^2\xi}{2,-2}+C\nm{\xi}{1,2,0}.
\label{D2X:est}
\end{equation}
Since $\nm{u}{k,p,\delta_1}\le \nm{u}{k,p,\delta_2}$ if
$\delta_1\ge\delta_2$, $\epsilon$ may be chosen such that
\bref{D2N:est}, \bref{D2X:est} combine to give
\begin{equation}
   \label{D2xi:est}
  \nm{\grado^2\xi}{2,-5/2}\le c(\nm{D\Phi_0(g,\pi)^*_1(\xi)}{2,-5/2} +
  \nm{D\Phi_0(g,\pi)^*_2(\xi)}{1,2,-3/2} ) +C\nm{\xi}{1,2,0},
\end{equation}
for smooth $\xi$.  Since $C_c^\infty$ is dense in $W^{2,2}_{-1/2}$, it
follows that \bref{D2xi:est} holds for all $\xi\in W^{2,2}_{-1/2}$.  Now
\bref{xi22} follows from the weighted Poincar\'e inequality \cite[Theorem
1.3]{Bartnik86}
\[
\nm{u}{p,\delta} \le c \nm{\grado u}{p,\delta-1} \le c \nm{\grado^2u}{p,\delta-2},
\]
for any $\delta <0$ and $u\in W^{2,p}_{\delta}$.
\qed


It will be useful to restructure $D\Phi^*$ into the operator $P^*$ defined
by
\begin{eqnarray}
\nonumber
   P^*(\xi) &=& \left[ \begin{array}{c}
    g^{1/4}\,(\nabla^{i}\nabla_jN -\delta^i_j\Lap_gN +(S^i_j-E^i_j)N)
    +g^{-1/4}\cL_X\pi^i_j 
\\[3pt]
   {} -g^{1/4}\nabla_p(2K^i_jN+\cL_Xg^i_j)  \end{array} \right]
\\
\label{P*}
&=& \rho \circ \left[ \begin{array}{cc} 1 & 0 \\ 0 & \nabla
\end{array}\right]
\circ D\Phi(g,\pi)^*\xi,
\end{eqnarray}
where $g^{1/4}=(\det g/\det \go)^{1/4}$ is a density of weight $\half$,
and 
\[
\rho = \rho(g) = \left[ 
\begin{array}{cc} g^{-1/4}g_{jk} & 0 \\ 0 & g^{1/4}g^{ik} 
\end{array}\right].
\]
The $L^2(dv_0)$-adjoint of $P^*$ is then
\begin{equation}
   P = D\Phi(g,\pi)\circ 
   \left[ \begin{array}{cc} 
          1 & 0 \\ 0 & -\delta_g \end{array}\right]
   \circ \rho,
\end{equation}
where $\delta_gq = \grad^p(q_p^{ij})$, so
$P(f_{i}^j,q_{pi}^{\ j})=D\Phi(f_{ij},-\grad^p(q_p^{\ ij}))$, and the
composition $PP^*$ is well-defined.

\begin{Proposition}
\label{P*:prop}
$P^*:W^{2,2}_{-1/2}(\cT)\to L^2_{-5/2}$ is bounded
and satisfies
\begin{equation}
   \label{P*est}
   \nm{\xi}{2,2,-1/2} \le c\nm{P^*\xi}{2,-5/2} +C\nm{\xi}{1,2,0},
\end{equation}
where $C$ depends on $\nm{(g,\pi)}{\cF}$, and $P^* = P^*_{(g,\pi)}$ has
Lipschitz dependence on $(g,\pi)\in\cF$,
\begin{equation}
   \label{P*Lip}
   \nm{(P^*_{(g,\pi)}-P^*_{(\tilde{g},\tilde{\pi})})\xi}{2,-5/2}
  \le C_1 \nm{(g-\tilde{g},\pi-\tilde{\pi})}{\cF}
        \nm{\xi}{2,2,-1/2},
\end{equation}
where $C_1$ depends on $\nm{(g,\pi)}{\cF}$,
$\nm{(\tilde{g},\tilde{\pi})}{\cF}$. 
\end{Proposition}

\Proof
That $P^*$ is bounded,
\begin{equation}
   \label{P*bnd}
   \nm{P^*_{(g,\pi)}\xi}{2,-3/2}
  \le C \nm{\xi}{2,2,-1/2},
\end{equation}
follows from estimates similar to but simpler than those of Proposition
\ref{DPhi*:coercive}.  The elliptic estimate \bref{P*est} follows directly
from \bref{xi22}.  $(P^*_{(g,\pi)}-P^*_{(\tilde{g},\tilde{\pi})})\xi$ is
controlled by breaking it up.  Since $\nm{g-\tilde{g}}{\infty}$,
$\nm{(N,X)}{\infty}$ are bounded by $\nm{g-\tilde{g}}{2,2,-1/2}$,
$\nm{\xi}{2,2,-1/2}$ respectively, terms such as
\[
   \left[ \begin{array}{cc} 
          g^{-1/4}-\tilde{g}^{-1/4} & 0 
          \\ 0 & g^{1/4}-\tilde{g}^{1/4} \end{array}\right]
   \circ
   \left[ \begin{array}{cc} 
          1 & 0 \\ 0 & -\delta_1 \end{array}\right]
   \circ
   D\Phi(g,\pi)^*\xi
\]
are controlled by $C\nm{g-\tilde{g}}{2,2,-1/2}\nm{\xi}{2,2,-1/2}$. Since
$\grad-\tilde{\grad} \sim \grado(g-\tilde{g})$, we may use \bref{u2:bnd} to
estimate, for example,
\[
\nm{(\grad-\tilde{\grad})D\Phi_2^*\xi}{2,-5/2}
\le c \nm{\grado(g-\tilde{g})}{1,2,-3/2}\nm{D\Phi_2^*\xi}{1,2,-3/2}.
\]
Using $D\Phi_2^*\xi = -2(NK_{ij}+\grad_{(i}X_{j)})$ shows
\begin{eqnarray*}
 \lefteqn{
\nm{D\Phi(g,\pi)_2^*\xi - D\Phi(\tilde{g},\tilde{\pi})_2^*\xi}{1,2,-3/2} }
\\&\le&
c \nm{N(K-\tilde{K})}{1,2,-3/2} + c\nm{\grado(g-\tilde{g})X}{1,2,-3/2},
\end{eqnarray*}
which is controlled by
\[
\nm{N}{\infty}\nm{K-\tilde{K}}{1,2,-3/2}+\nm{\grado N(K-\tilde{K})}{2,-5/2}
\]
for the first, and similarly for the second term.  Again using the
$L^\infty$ bound and \bref{u2:bnd} controls the difference by
$C\nm{\xi}{2,2,-5/2}$ as required; the terms in $ D\Phi(g,\pi)_1^*\xi -
D\Phi(\tilde{g},\tilde{\pi})_1^*\xi $ are controlled by very similar
estimates, giving \bref{P*Lip}.
\qed

We now show that the elliptic estimate is  also satisfied by weak solutions,
which are \emph{a priori} only in $L^2$.  We say that $\xi\in\cL$ is a
\emph{weak solution} of $D\Phi(g,\pi)^*(\xi)=(f_1,f_2)$ for $(f_1,f_2)\in
L^2_{-3/2}(\tilde\cS)\times W^{1,2}_{-3/2}(\cS)$ if
\begin{equation}
\label{wDPhi*}
   \int_\cM \xi \cdot D\Phi(g,\pi)(h,p) 
    = \int_\cM (f_1,f_2)\cdot(h,p),\quad \forall\ (h,p)\in \cG\times\cK.
\end{equation}
In this definition it suffices to test with just $(h,p)\in 
C_c^\infty(\cS\times\cSt)$, since this space is dense in $\cG\times\cK$.

\begin{Proposition}\label{ker-reg:ppn}
Suppose $(g,\pi)\in\cF$, $(f_1,f_2)\in L^2_{-3/2}(\tilde\cS)\times
W^{1,2}_{-3/2}(\cS)$, and $\xi=(N,X^i)\in\cL = L^2_{-1/2}(\cT)$ is a weak
solution of $D\Phi(g,\pi)^*(\xi)=(f_1,f_2)$.  Then $\xi\in
W^{2,2}_{-1/2}(\cT)$ is a strong solution and $\xi$ satisfies \bref{xi22}.
\end{Proposition}

\Proof
We first show $\xi\in W^{2,2}_\loc$, so restrict to a coordinate
neighbourhood $\Omega$. In local coordinates $P^*(\xi)=f$ is equivalent to
relations of the form
\[ 
  A\cdot\partial^2\xi+B\cdot\partial\xi + C\xi =f,
\]
where $A:\bR^{36}\to\bR^{36}$ is invertible and determined solely by $g$;
see \bref{D2N}, \bref{Xijk:idn}.  Furthermore, $A\in W^{2,2}$, $B\in
W^{1,2}$, $C\in L^2$ in $\Omega$, so this is equivalent to 
\begin{equation}
   \partial^2_{ij}\xi^\alpha +\partial_k(b^{k\alpha}_{ij\beta}\xi^\beta)
 + c_{ij\beta}^\alpha \xi^\beta = f^{\alpha}_{ij}
\label{D2xi}
\end{equation}
for suitable $b\in W^{1,2}$, $c,f\in L^2$.  Thus $\xi\in L^2$ satisfies the
weak form of \bref{D2xi},
\[
  \int_\Omega (\partial^2_{ij}\phi^{ij}_\alpha +
  b_{ij\alpha}^{k\beta}\partial_k\phi_{\beta}^{ij} +
  c_{ij\alpha}^{\beta}\phi_\beta^{ij} )\xi^\alpha\, dx
  = \int_\Omega \phi^{ij}_\alpha f_{ij}^\alpha \, dx,
\]
for all $\phi\in W_c^{2,2}(\Omega)$.
Replacing $\phi$ by $J_\epsilon \phi$ where $J_\epsilon$ is a Friedrichs
mollifier with mollification parameter $\epsilon>0$, we see that
$\xi_\epsilon =J_\epsilon\xi$ is smooth and satisfies
\[
\partial^2 \xi_\epsilon + \partial J_\epsilon(b\xi) + J_\epsilon(c\xi) =
J_\epsilon f.
\]
Following a suggestion of L.~Simon, we let $u=\chi\xi_\epsilon$ where $\chi\in
C_c^\infty(\Omega)$ is any cutoff function.  Then taking a trace shows that
$u\in C_c^\infty(\Omega)$ satisfies an equation of the form
\[ 
\Lap_0 u= F+\partial G,
\]
where $F=F_1+F_2+F_3$, $G=G_1+G_2$ and $F_1=\chi''\xi_\epsilon+\chi
J_\epsilon f$, $F_2=\chi' J_\epsilon(b\xi)$, $F_3=\chi J_\epsilon(c\xi)$,
$G_1=\chi'\xi_\epsilon$, $G_2=\chi J_\epsilon(b\xi)$.  The terms $F,G$ are
smooth with compact support, so $u$ has a representation
\[
u(x) = \Gamma * (F+\partial G) = \int_\Omega \Gamma(x-y)(F(y)+\partial
G(y))\,dy,
\]
where $\Gamma(x-y) = (4\pi|x-y|)^{-1}$ is the fundamental solution of
Laplace's equation.  Let $D=(-\Lap_0)^{1/2}$ be the Riesz potential
\cite[Ch.~V]{Stein70}.  The operators $K_{ij}=\partial^2_{ij}\Gamma$ and
$K_i=\partial_i \Gamma D$ are Calderon-Zygmund kernels in the sense of
\cite[Ch.~II]{Stein70}, and hence satisfy
\begin{equation}
   \nm{K_{ij} * w}{L^p(\Omega)}+ \nm{K_{i} * w}{L^p(\Omega)}\le
   c\nm{w}{L^p(\Omega)}  .
\label{CKij}
\end{equation}
We now use these bounds to control the various terms in $\Gamma *
(F+\partial G) $ and thereby bootstrap the estimates for $u$ up to a
$W^{2,2}$ bound which is independent of $\epsilon$.

Since $K_{ij}*F_1=\partial_{ij}^2(\Gamma *F_1)$,  \bref{CKij}
with $p=2$ shows that 
\[
\nm{\Gamma *F_1}{2,2} \le c\nm{F_1}{2} \le c(\nm{\xi}{2}+\nm{f}{2}),
\]
where the norms here are over $\Omega$. In particular, $\Gamma* F_1$ is
uniformly bounded in $W^{2,2}(\Omega)$, independent of $\epsilon$.  Since
$b\xi\in W^{1,2}\cdot L^2 \subset L^6\cdot L^2 \subset L^{3/2}$, $F_2$ is
uniformly bounded in $L^{3/2}$ and thus
\[
  \nm{\Gamma*F_2}{2,3/2} \le c\nm{F_2}{3/2} \le c\nm{b}{1,2}\nm{\xi}{2}.
\]
Now $F_3\in L^1(\Omega)$ only, so we instead note that $\partial_i
u=K_i*(Du)$ where $D$ satisfies 
\[
\nm{Dw}{p}\le c\nm{w}{q},
\]
for either $1<q<n$ with $1/p=1/q-1$ , or if $q=1$, with any
$1<p<n/(n-1)=3/2$.  With $q=1$ and $p<3/2$ we thus have
\[
\nm{\partial\Gamma*F_3}{p} \le c\nm{D*F_3}{p}\le c\nm{F_3}{1},
\]
and the Sobolev inequality now shows that $\nm{\Gamma*F_3}{3-\delta}$ is
uniformly bounded in terms of $\nm{c}{2}\nm{\xi}{2}$, for any small
$\delta>0$.  Now $\nm{G_1}{2}\le c\nm{\xi}{2}$, so we use the identity
$\Gamma*(\partial_k G_1^k) = \partial_k\Gamma *G^k_1$ and the Sobolev inequality
to estimate 
\[
\nm{\Gamma*(\partial G_1)}{6} \le c\nm{K_k*G^k_1}{2} \le c\nm{G_1}{2} \le
c\nm{\xi}{2}.
\]
Likewise, since $G_2\in L^{3/2}$ uniformly, we find by a similar argument
that $\nm{\Gamma*(\partial g_2)}{3} \le c\nm{b}{1,2}\nm{\xi}{2}$.

Assembling all the pieces now shows that $\xi\in L^{3-\delta}_\loc$, for
any $\delta>0$, and we now repeat the above arguments with this stronger
bound on $\xi$.  Bootstrapping in this way shows eventually that $\xi\in
W^{2,2}_\loc$.  Thus $\chi_R\xi\in W^{2,2}_{-1/2}$ for any cutoff function
$\chi_R\in C_c^\infty(\cM)$, $\chi_R(x)=\chi(x/R)$ with $\chi_R(x)=1$ on
$B_R$.  Now \bref{xi22} shows that $\chi_R \xi$ is uniformly bounded in
$W^{2,2}_{-1/2}$ since $\xi\in L^2_{-1/2}$ and $\chi_R\xi \to \xi$, so
$\xi\in W^{2,2}_{-1/2}$ as required.
\qed


We next show that the kernel of $D\Phi^*$ is trivial in the space of
lapse-shift pairs decaying at infinity.  We may interpret this result
as saying there are no generalised Killing vectors decaying to zero at
infinity, where by {\em generalised Killing vector\/} $\xi$ of
$(g,\pi)\in\cF$, we mean that $\xi\in W^{2,2}_{\loc}(\cT)$ satisfies
$D\Phi(g,\pi)^*\xi=0$.  Likewise, if there exists a nontrivial vector
field $\xi$ satisfying $D\Phi(g,\pi)^*\xi=0$ then $(g,\pi)$ is a
\emph{Killing initial data} set, where the terminology is motivated by
a result of Moncrief \cite{Moncrief76} which shows that if $(g,\pi)$
 satisfies the constraint equations, then a generalised
Killing vector determines a standard Killing vector field in the
spacetime generated from the initial data $(g,\pi)$ by solving the
vacuum Einstein equations.  Of course, this requires that $(g,\pi)$
has enough regularity that a local existence and uniqueness theorem
for the Einstein evolution can be applied, which is not the case at
present for general $(g,\pi)\in\cF$.  However, if local existence and
uniqueness could be established for $s=2$ then it would be possible to
identify a generalised Killing vector (ie.~$\xi\in\ker D\Phi^*$) with
the spatial restriction of a true vacuum spacetime Killing vector.

\begin{Theorem}  \label{kerDPhi=0:thm}
        Suppose $\Omega\subset\cM$ is a connected domain and $E_R \subset
\Omega$ for some exterior domain $E_R$, fix $(g,\pi)\in \cF$ and suppose
$\xi\in L^2_{-1/2}(\cT)$ satisfies $D\Phi(g,\pi)^*\xi = 0$ in $\Omega$.
Then $\xi\equiv0$ in $\Omega$.
\end{Theorem}

\Proof By Proposition \ref{ker-reg:ppn}, $\xi\in W^{2,2}_{-1/2}(\cT)$ and
\bref{Xijk:idn}, the equation $D\Phi(g,\pi)^*\xi = 0$ shows that $\xi$
satisfies an equation of the form
\begin{equation}
        \grado^2\xi = b_1\grad\xi + b_0 \xi,
        \label{D2phi}
\end{equation}
with coefficients $b_0 \in L^2_{-5/2}$, $b_1\in W^{1,2}_{-3/2}$.  We must
now show that a solution of \bref{D2phi} which decays as $\xi=o(r^{-1/2})$,
must vanish.  The structure of the argument to follow is well-known: the
difficulty here lies in the absence of the continuity assumptions used
essentially in \cite{ChristodoulouOMurchadha81}.

If $u\in W^{1,2}_0(\bR^n)$ then the Sobolev inequality is true in the sharp
form $\nm{u}{n/(n-1)}\le c \nm{Du}{1}$.  Such an inequality remains valid
without the hypothesis of compact support, provided $u$ vanishes on a
sufficiently large set.
\begin{Lemma}
  \label{sobolev:lem}
  Suppose $n\ge3$, $B_R\subset\bR^n$, $1\le p <\infty$, and $q\le
  np/(n-p)$ if $p<n$, $q<\infty$ if $p=n$, $q\le \infty$ if $p> n$.
  If $u\in W^{1,p}(B_R)$ satisfies $u\equiv0$ in $B_{\eta R}$ for some
  $0<\eta\le1$, then
    \begin{equation}
        \nm{u}{q;B_R} \le c \eta^{2(1-n/p)}R^{1+n/q-n/p}\nm{Du}{p;B_R}.
    \end{equation}
\end{Lemma}
\Proof By rescaling we may assume $R=1$.  Let $\tilde{u}(x) = u(\psi(x))$,
where $\psi:\bR^n\backslash\{0\} \rightarrow \bR^n\backslash\{0\}$ is the
inversion map, $\psi(x)=x/\len{x}^2$.  Since $1\le \len{d\psi(x)}\le
\eta^{-2}$ for $\eta\le\len{x}\le1$, we see that $\tilde{u}\in
W^{1,p}(\bR^n\backslash B_1)$ and $\tilde{u}(x)=0$ for
$\len{x}\ge\eta^{-1}$.  The usual argument for the Sobolev inequality in
$\bR^n$ applies also to $\bR^n\backslash B_1$ (see \cite[Chapter
7]{GilbargTrudinger77}) and shows that for $p<n$,
\begin{displaymath}
    \nm{\tilde{u}}{np/(n-p)} \le c \nm{D\tilde{u}}{p};
\end{displaymath}
it is not necessary that $\tilde{u}$ be defined in $B_1$.  Now
$\nm{\tilde{u}}{q}\le c\nm{\tilde{u}}{np/(n-p)}$ gives
\[
\nm{\tilde{u}}{q} \le c \nm{D\tilde{u}}{p}; 
\]
if $p\ge n$ then this estimate follows similarly from Sobolev embedding.
The result now follows from the bounds $1\le|d\psi(x)|\le\eta^{-2}$
and rescaling.
\qed
\begin{Lemma}\label{D2u=0:lem}
        Suppose $\Omega \subset \bR^3$ and $u = (u^1,\dots,u^K) \in 
        W^{2,2}(\Omega,\bR^K)$ satisfies
        \begin{equation}
                D^2_{ij}u^A = a^{AB}_{ij}u^B + b^{AB}_{ijk}D_ku^B,
                \label{D2u:eqn}
        \end{equation}
        where $a\in L^2(\Omega,\bR^{9K^2})$, $b\in L^6(\Omega,\bR^{27K^2})$.
        Then there is a constant $R_1>0$, depending on $\nm{a}{2}, \nm{b}{6}$,
        such that if $R\le R_1$, $B_R(x_0)\subset\Omega$,
        and $u\equiv0$ in $B_{R/2}(x_0)$, then $u\equiv0$ in $B_R(x_0)$.
\end{Lemma}
\Proof
Since $u=0$ in $B_{R/2}$, Lemma \ref{sobolev:lem} may be applied with
$q=\infty$ and $q=6$ to give
\begin{eqnarray}
        \nm{D^2u}{2;B_R} & \le & \nm{a}{2;\Omega} \nm{u}{\infty;B_R}
                          +\nm{b}{6;\Omega}\nm{Du}{3/2;B_R}
        \nonumber \\
         & \le & c R^{1/2} \nm{a}{2;\Omega} \nm{D^2u}{2;B_R}
                +c R^{3/2} \nm{b}{6;\Omega} \nm{Du}{6;B_R}
        \nonumber \\
         & \le & c R^{1/2}\left(\nm{a}{2;\Omega}+R\nm{b}{6;\Omega}\right)
                 \nm{D^2u}{2;B_R}.
\end{eqnarray}
Thus if 
$R\le R_1 = \half \min\{1,c^{-2}(\nm{a}{2;\Omega}+\nm{b}{6;\Omega})^{-2}\}$ 
then $\nm{D^2u}{2;B_R}=0$ and hence $u\equiv 0$ as claimed.
\qed

\begin{Proposition} \label{kerlocal:ppn}
        Suppose $(g,\pi)\in\cF$, $\xi=(N,X^i)$ satisfies $D\Phi(g,\pi)^*\xi=0$ 
        in a connected subset
        $\Omega\subset\cM$, and $\xi\equiv0$ in 
        some open set $U\subset\Omega$.  Then $\xi\equiv0$ in $\Omega$.
\end{Proposition}
\Proof
   We may cover $\Omega\subset\cM$ by a finite set of coordinate neighbourhoods in 
   which $C^{-1}\len{v}^2 \le g_{ij}v^iv^j \le c \len{v}^2, \ \forall 
   \ v = v^i\del_i$, where $\len{v}^2 = \Sigma(v^i)^2$.
   Since $\grad^2_{ij} = D^2_{ij} - \Gamma_{ij}^k D_k$, after moving 
   some Christoffel terms into $b_1$ the equation \bref{D2phi} in a given 
   coordinate chart $\Omega'$ may be written symbolically as 
   \begin{equation}
                  D^2\xi = a \xi + b D\xi,
                  \label{lapse:eqn}
   \end{equation}
   where $a\in L^2(\Omega')$, $b\in W^{1,2}(\Omega')\subset L^6(\Omega')$
   and 
   $$
   \nm{a}{2;\Omega'}+R\nm{b}{6;\Omega'} 
     \le C \left(\nm{g-\go}{2,2-1/2} + \nm{\pi}{1,2,-3/2}\right).
   $$
   We can apply the previous lemma in each coordinate chart: in particular,
   if $\xi\equiv0$ in some open set $U\subset\Omega$ but $\xi\ne0$ at some
   point of $\Omega$, then there is a coordinate chart $\Omega'$ and a ball
   $B_{R_2}(x_0)\subset\Omega'$ such that $\xi\equiv0$ in $B_{R_2}(x_0)$
   but $\xi\not\equiv0$ in $B_{R_3}(x_0)$ for every $R_3\ge R_2$.  But
   Lemma \ref{D2u=0:lem}, applied to $B_{R/2}(x_0+(R_2-R/2)e)$ for any unit
   vector $e\in \bR^3$ and $R\le R_1$, shows that $\xi\equiv0$ in
   $B_{R_2+R/2}(x_0)$, which is a contradiction. Thus $\xi$ vanishes in the
   coordinate set $\Omega'$, and hence in all $\Omega$ since it is
   connected.  \qed
   
   To complete the proof of Theorem \ref{kerDPhi=0:thm} we must show
   $\xi$ vanishes near infinity.  To do this we establish a weighted
   Poincar\'e inequality about the point at infinity.

\begin{Lemma}
        \label{Du-inf:lem}
        Suppose $p,\delta$ satisfy $p\ge1$, $\len{\delta p/n + 1} < 1$ and 
        $u\in W^{1,p}_\delta(E_R)$, $E_R\subset \bR^n$,
         then there is $c=c(n,p,\delta)$ such that
        \begin{equation}
                \nm{u}{p,\delta;E_R} \le c \nm{Du}{p,\delta-1;E_R}.
                \label{Du-inf:est}
        \end{equation}
\end{Lemma}

\Proof Since $C^\infty_c(E_R)$ is dense in $W^{1,p}_\delta(E_R)$, it
suffices to prove \bref{Du-inf:est} for smooth, compactly supported $u$.
For $\lambda\in\bR^+, f\in C^\infty_c(E_R)$, ${d\over d\lambda} f(\lambda
x) = |x| D_r f(\lambda x)$ implies
$$
f(x) = - \int_1^\infty \len{x} D_r f(\lambda x)\, d\lambda,
$$
because $f(x)=0$ for $r=|x|$ sufficiently large.  Hence
\begin{eqnarray*}
        \int_{E_R} \len{f(x)}\,dx & \le & 
                \int_{1}^{\infty}\int_{E_R}\len{x} \len{D_r f(\lambda x)}
                                     \,dx\,d\lambda  \\
         & \le & \int_{1}^{\infty}\int_{E_{\lambda R}}\len{x} \len{D_r f(x)}
                          \,dx\,\lambda^{-n-1}d\lambda  \\
         & \le & {1\over n} \int_{E_R} \len{x} \len{D_rf(x)} \,dx.
\end{eqnarray*}
Now substituting $f(x) = |u(x)|^{p}|x|^{-\delta p - n}$, whence
$$
\len{D_rf} \le p \len{u}^{p-1} \len{Du} \len{x}^{-\delta p - n} 
    + \len{\delta p+n} \len{u}^p \len{x}^{-\delta p -n-1},
$$
we find that 
\begin{eqnarray*}
\lefteqn{ {1\over n} \int_{E_R}\len{x}\len{D_rf}\,dx  \le  
              \len{1+\delta p/n} \int_{E_R} \len{u}^p \len{x}^{-\delta p-n}dx   } 
              \qquad{}\qquad\\
         &&{} + {p\over n}
           \left(\int_{E_R}\len{u}^p \len{x}^{-\delta p-n}dx\right)^{1-1/p}
           \left(\int_{E_R}\len{Du}^p \len{x}^{-(\delta-1) p-n}dx\right)^{1/p}.    
\end{eqnarray*}
Thus if $\len{1+\delta p/n} < 1$, then
\begin{equation}
        \nm{u}{p,\delta;E_R} \le {p/n \over 1-\len{1+\delta p/n}} 
        \nm{Du}{p,\delta-1;E_R},
\end{equation}
as required.
\qed

In particular, in $\bR^3$ and in $E_R\subset\cM$ we have the estimates
\begin{eqnarray}
        \nm{Du}{2,-3/2;E_R} & \le & \tfrac{2}{3} \nm{D^2u}{2,-5/2;E_R},
        \label{Du1} \\
        \nm{u}{2,-1/2;E_R} & \le & 2 \nm{Du}{2,-3/2;E_R},
        \label{uDu2}
\end{eqnarray}
for $R\ge R_0$, valid whenever both sides of the inequalities are finite.
Using the weighted H\"older and Sobolev inequalities \cite{Bartnik86},
we have in $E_R$,
$$
\nm{D^2\xi}{2,-5/2} \le \left(\nm{b_0}{2,-5/2}\nm{\xi}{\infty}
        + c\nm{b_1}{6,-3/2}\nm{D\xi}{3,-1}\right).
$$
But from \bref{Du1},
\begin{eqnarray*}
        \nm{D\xi}{3,-1} & \le & \nm{D\xi}{1,2,-1}  \\
         & \le & R^{-1/2} \left(\nm{D\xi}{2,-3/2}+ \nm{D^2\xi}{2,-5/2}\right)  \\
         & \le & c R^{-1/2}\nm{D^2\xi}{2,-5/2},
\end{eqnarray*}
since $\nm{u}{p,\delta;E_R}\le R^{\eta-\delta}\nm{u}{p,\eta;E_R}$
for $\eta < \delta$.
Thus there is $R_1 = R_1(\nm{b_1}{1,2,-3/2})$ such that for any $R\ge R_1$ we have
\begin{equation}
        \nm{D^2\xi}{2,-5/2;E_R} \le c \nm{b_0}{2,-5/2} \nm{\xi}{\infty,0;E_R}.
        \label{ccc}
\end{equation}
Since for any $u\in W^{2,2}_{-1/2}$,
\begin{eqnarray*}
        \nm{u}{\infty,0;E_R} & \le & R^{-1/2}\nm{u}{\infty,-1/2;E_R} \\
         & \le & c R^{-1/2}\nm{u}{2,2,-1/2;E_R}  \\
         & \le & c R^{-1/2}\nm{D^2u}{2,-5/2,E_R},
\end{eqnarray*}
by the Sobolev inequality and Lemma \ref{Du-inf:lem}, it follows that
\begin{equation}
        \nm{D^2\xi}{2,-5/2;E_R} \le C R^{-1/2}\nm{D^2\xi}{2,-5/2;E_R}
\end{equation}
and thus $\xi$ vanishes in $E_R$ for $R$ sufficiently large.
Combining this result with Proposition \ref{kerlocal:ppn}
 completes the proof of Theorem \ref{kerDPhi=0:thm}, since $\Omega\subset \cM$ 
is assumed to be connected.
\qed

\begin{Corollary}
\label{P*:cor}
There is a constant $C_2$ depending on $\nm{(g,\pi)}{\cF}$ such that
for all $\xi\in W^{2,2}_{-1/2}$,
\begin{equation}
   \nm{\xi}{2,2,-1/2} \le C_2 \nm{P^*\xi}{2,-5/2}.
\label{P*sharp}
\end{equation}
\end{Corollary}
\Proof This follows from a standard Morrey contradiction argument.  Suppose
not, so there is a sequence $\xi_k$, $k=1,2,\dots$ such that
$\nm{\xi_k}{2,2,-1/2} =1$, $\nm{P^*\xi_k}{2,-5/2}\le 1/k$.  Then
$P^*\xi_k\to0$ strongly in $L^2_{-5/2}$.  Now $W^{2,2}_{-1/2}$ embeds
compactly in $W^{1,2}_0$, so $\xi_k$ converges strongly in $W^{1,2}_0$, to
$\xi$ say.  Applying \bref{P*est} to $\xi_j-\xi_k$ shows that $\xi_k$ is a
Cauchy sequence in $W^{2,2}_{-1/2}$ and hence converges strongly to $\xi$
in $W^{2,2}_{-1/2}$.  Then $\nm{\xi}{2,2,-1/2}=1$ and $P^*\xi=0$, which
contradicts the triviality of $\ker P^*$ (Theorem \ref{kerDPhi=0:thm}).
\qed

The Implicit Function Theorem method is used  to conclude that
$\cC$ is a smooth Hilbert submanifold of $\cF$ --- in fact we show that
{\em all\/} level sets of $\Phi$ are smooth submanifolds.
\begin{Theorem}
\label{constraintmfld:thm}
For each $(\varepsilon,S_i)\in \cL^*$,  the constraint set
\begin{equation}
\cC(\varepsilon,S_i) = \{ (g,\pi)\in\cF : \Phi(g,\pi) = 
                          (\varepsilon,S_i) \}
\label{CeS:mfld}
\end{equation}
is a Hilbert submanifold of $\cF$.  In particular, the space of solutions
of the vacuum constraint equations, $\cC=\Phi^{-1}(0) = \cC(0,0)$, is a
Hilbert manifold.
\end{Theorem}
\Proof By the implicit function theorem, it suffices to show that
$D\Phi:\cG\times\cK \rightarrow \cL^*$ is surjective and splits.
Since $D\Phi$ is bounded, its kernel is closed and hence splits.  We
have shown in Proposition \ref{ker-reg:ppn} and Theorem
\ref{kerDPhi=0:thm} that $\ker\{ D\Phi(g,\pi)^*:\cL \rightarrow
(\cG\times\cK)^*\} = \{0\}$, so the cokernel of $D\Phi$ is trivial.
It remains to show that $D\Phi$ has closed range, which we show by a
direct argument.  Note that the argument of Fischer-Marsden
\cite{FischerMarsden79} based on the ellipticity of $PP^*$ encounters
some difficulties, arising from the low regularity of some low order
coefficients (such as $\nabla^2 \Ric$) of $PP^*$, and we have not been
able to overcome these problems.  This difficulty appears to restrict
the Fischer-Marsden elliptic method to neighbourhoods of data $(g,\pi)$
which are 2 derivatives smoother, ie. $H^4\times H^3$.

Instead we consider particular variations $(h,p)$ of $(g,\pi)$
determined from fields $(y,Y^i)$, of the form (cf.~\cite{CorvinoSchoen03})
\[
h_{ij} = 2 y g_{ij},\quad p^{ij} = (\nabla^i Y^j+\nabla^jY^i -
\grad_kY^k g^{ij})\rtg
\]
and define 
\begin{eqnarray}
\lefteqn{ F(y,Y) = D\Phi(h,p)}
\nonumber \\
&=& \left[ 
  \begin{array}{c}
    -4\rtg\Lap y +\Phi_0(g,\pi)y + \trg\pi \grad_kY^k
     -4\pi\sbullet\grad Y
\\
  2\rtg(\Lap Y_i + \Ric_{ij}Y^j) + 2\Phi_i(g,\pi)y +
    (4\pi_i^j-2\trg\pi\delta_i^j)\grad_jy 
  \end{array}
\right].
\label{F}  
\end{eqnarray}
We see that if $y\in W^{2,2}_{-1/2}(\cM)$, $Y\in W^{2,2}_{-1/2}(T\cM)$
then $(h,p)\in \cG\times\cK$, and it is straightforward to check that
\begin{equation}
  \label{Fbnd}
  F:  W^{2,2}_{-1/2}(\cM)\times W^{2,2}_{-1/2}(T\cM) \to
  L^2_{-5/2}(\cT^*\otimes\Lambda^3) = \cL^*
\end{equation}
is bounded.  Moreover, the general scale-broken elliptic estimate \cite{Bartnik86}
\[
\nm{u}{2,2,-1/2} \le c \nm{\Lap u}{2,-5/2} + C\nm{u}{2,0}
\]
shows that
\begin{eqnarray*}
\nm{(y,Y)}{2,2,-1/2} &\le& c\nm{F(y,Y)}{2,-5/2} + C\nm{(y,Y)}{2,0}
\\
&&  \nm{\Phi y }{2,-5/2}  + \nm{\pi\grad(y,Y)}{2,-5/2} 
  + \nm{\Ric(Y)}{2,-5/2},
\end{eqnarray*}
and the last terms are estimated by H\"older, Sobolev and
interpolation inequalities, eg:
\begin{eqnarray*}
  \nm{\pi \grad u}{2,-5/2}^2 &\le& c\nm{\grad u}{3,-1}\nm{\pi}{6,-3/2}
\\
&\le& c \nm{\pi}{1,2,-3/2} \nm{\grad u}{3,-1}
\\
&\le& \epsilon \nm{u}{2,2,-1/2} +C \nm{u}{2,0},
\end{eqnarray*}
where $C$ depends on $\epsilon$, $\lambda$ and $\nm{(g,\pi)}{\cF}$ as
usual. Thus $F$ satisfies the scale-broken estimate
\begin{equation}
  \label{Fest}
  \nm{(y,Y)}{2,2,-1/2} \le c \nm{F(y,Y)}{2,-5/2} + C\nm{(y,Y)}{2,0}.
\end{equation}
Now the adjoint $F^*$ has a similar structure and the same argument
shows $F^*$ also satisfies an estimate \bref{Fest}.  It follows that
$F$ has closed range (from \bref{Fest}) with  finite dimensional
cokernel (since $F^*$ has finite dimensional kernel by the elliptic
estimate for $F^*$).   Since clearly $\mathrm{ran}\, F \subset
\mathrm{ran}\,D\Phi$, we have shown that $D\Phi$ has closed range and
the proof of Theorem \ref{constraintmfld:thm} is complete.
\qed

\section{ADM energy-momentum}\label{ADM:sec}
%
%

The ADM total energy-momentum $\bPadm(g,\pi) = (\bPadm_\alpha) = (E,p_i)$ 
is usually defined by the formal expressions
\begin{eqnarray}
        16\pi E  & =  & \oint_{S_{\infty}} \left(\partial_ig_{ij} - 
        \partial_jg_{ii}\right)\,dS^j
        \label{E-formal:def} \\
        16\pi p_i & = & 2 \oint_{S_{\infty}} \pi_{ij}\, dS^j
        \label{p-formal:def}
\end{eqnarray}
where $dS^j$ is the normal element of the sphere at infinity
$S_{\infty}$, the indices refer to a suitable rectangular coordinate
system near infinity, and the integral over $S_\infty$ is understood
as a limit of integrals over finite coordinate spheres.  The
expression for the total energy $E$ was investigated in
\cite{Bartnik86} and shown to be well-defined (that is, independent of
the limiting process used to define ${S_{\infty}}$ and of the choice
of structure at infinity), for metrics satisfying $g-\go \in
W^{2,q}_{-1/2}$ for some $q>3$, and $R(g)\in L^1$.  In this section we
reformulate \bref{E-formal:def}, \bref{p-formal:def} and show that the
redefined $\bPadm$ is well-defined under weaker regularity conditions,
which are better adapted to the Hilbert manifold structure of $\cC$.
It is not immediately clear the formal definitions
\bref{E-formal:def}, \bref{p-formal:def} can be made sensible under
the weaker conditions; that this can be done, with result agreeing
with the definitions \bref{Eadm:def}, \bref{padm:def} below, is shown
in Proposition \ref{ADMdef-ok:pro}.  We also show that $\bPadm$ is
independent of the choice of structure of infinity, thereby extending
the mass uniqueness result of \cite{Bartnik86}.

The first result implies in particular that $\bPadm$ (after suitable
reformulation) defines a bounded function from the (vacuum) constraint
manifold $\cC$ to $\bR^4$, which is smooth with respect to the Hilbert
manifold structure of $\cC$.  However, it turns out that the definition of
$\bPadm$ cannot be extended to all $(g,\pi)\in\cF$ as a bounded
(well-defined) function.  This restriction is not an artifact of the rather
weak regularity conditions of $\cF$; rather it reflects the need for
additional decay conditions in defining $\bPadm$.  In the usual physics
framework, where $(g,\pi)$ satisfy the decay conditions (with $r=\len{x}$),
\begin{eqnarray}
        \len{g_{ij}-\delta_{ij}} +  r \len{\partial_ig_{jk}} 
           + r^2\len{\partial_i\partial_jg_{kl}}  & =  & O(1/r),
        \label{d2g-phys:dcy} \\
        \len{K_{ij}} + r\len{\partial_iK_{jk}} & =  & O(1/r^2),
        \label{dK-phys:dcy}
\end{eqnarray}
the nature of the additional decay conditions is usually expressed by the 
requirements \cite{York80}
\begin{equation}
        R(g) = O(r^{-4}),\qquad \partial_iK_{ij} - \partial_jK_{ii} = O(r^{-4}).
        \label{L1-phys:dcy}
\end{equation}
These may be reformulated more invariantly (and more generally) as
\begin{equation}
        R(g) \in L^1(\cM),\qquad \grad_j \pi^{ij}\in L^1(T\cM),
        \label{L1-geom:dcy}
\end{equation}
and we emphasise that these conditions are {\em not\/} satisfied by
general $(g,\pi)\in\cF$.  Indeed, they are equivalent to requiring
$\Phi(g,\pi)\in L^1(\cT^*)$, and exactly this condition turns out to
be sufficient for $\bPadm(g,\pi)$ to be well-defined.

In order to define $\bPadm$ in all of $\cF$, we first need a suitable
definition of translation vector at infinity.  Fix a 4-vector
$\xinf=(\xinf^\alpha)=(\xinf^0,\xinf^i)\in\bR^4$ (where the indices
take the ranges $\alpha=0,1,\dots,3,\ i=1,\dots,3$); using the metric
$\go$ near infinity, which we consider as defining a connection on the
spacetime tangent bundle $\cT$ which is flat near infinity, we may
identify $\xinf$ with a parallel vector field $\tilde{\xi}_\infty$
defined in an exterior region $E_{R_1}$ for some $R_1\ge R_0$.  We say
that a vector field $\xhinf\in C^{\infty}(\cT)$ is a {\em constant
  translation near infinity\/} representing $\xinf \in\bR^4$ if
$\xhinf = \tilde{\xi}_\infty$ in $E_{2R_1}$ and $\xhinf = 0$ in
$\cM\setminus E_{R_1}$.  Obviously $\xhinf$ is not uniquely determined
by its constant value $\xinf$; however two representatives of $\xinf$
differ only by a smooth, compactly supported, vector field.

A vector field $\xi=(\xi^\alpha)$ is then said to be an {\em
  asymptotic translation\/} if there is $\xinf\in\bR^4$ with a
corresponding constant translation vector at infinity $\xhinf$, such
that $\xi - \xhinf\in L^{2}_{-1/2}(\cT) = \cL$.  Note that if
$\xi^{(1)}, \xi^{(2)}$ are two asymptotic translations (representing
the same translation vector $\xinf$), then $\xi^{(1)}-\xi^{(2)}
\in\cL$; hence we may define the class
\begin{equation}
        \xinf + \cL = \{\xi : \xi-\xhinf \in \cL \},
        \label{xinf+L:def}
\end{equation}
of asymptotic translation vector fields representing $\xinf$.
By replacing $\cL$ with $W^{k,2}_{-1/2}(\cT)$, $k\ge1$, we may similarly 
define classes of asymptotically constant vectors with better regularity 
properties.

Rather than work with the asymptotic boundary integrals 
\bref{E-formal:def}, \bref{p-formal:def}, it is more convenient
(although logically equivalent, as we shall show) 
to work with spatial integrals of exact 
divergences.  Therefore we introduce the density-valued linear operators 
$\cRo(g), \cPo(\pi)$ by 
\begin{eqnarray}
        \cRo(g)     &=& \biggl(\grado^{ij}g_{ij} - \Lapo\tro g\biggr)\rto, 
\label{Ro:def}   \\
        \cPo{}_i(\pi) &=& \go{}_{ij}\grado_k \pi^{jk} .
\label{Po:def}
\end{eqnarray}

The ADM (total) energy-momentum vector $\bPadm(g,\pi)=(E,p)$ is then 
defined by describing the pairing with a vector at 
infinity $\xinf\in\bR^{3,1}$; let $\xhinf$ be a corresponding 
representative translation vector field at infinity, 
then we define $\xinf^\alpha \bPadm_\alpha(g,\pi)$ by
\begin{eqnarray}
        16\pi \xi_\infty^0 \bPadm_0(g,\pi) & = & \int_{\cM}
           \left(\xhinf^0\cRo(g) + \grado^i\xhinf^0 
              \left(\grado^jg_{ij}-\grado_i \tro g\right)\rto,
           \right)
        \label{Eadm:def} \\*
        16\pi \xi_\infty^i \bPadm_i(g,\pi) & = & 2\int_{\cM}
           \left(\xhinf^i \cPo_i(\pi) +  \pi^{ij} \grado_i\xhinf{}_j\right),
        \label{padm:def}
\end{eqnarray}
where indices are raised and lowered using the background metric $\go$.
The physical interpretation of $\xi^\alpha\bPadm_\alpha$ is as the energy of 
$(\cM,g,\pi)$ as observed by the asymptotic time vector $\xinf$, and 
$\bPadm_\alpha$ is the total energy-momentum covector of $(\cM,g,\pi)$.
Since $\cRo(g), \cPo(\pi), \rto, \pi$ are all tensor densities, 
the volume elements in \bref{Eadm:def},\bref{padm:def} are present 
implicitly, and it is readily seen that the right hand sides depend only 
on $\xinf$ and not on the specific choice of representative asymptotic 
translation vector $\xhinf$, since a change of $\xhinf$ changes the integrands 
only by an exact divergence of compact support.

\begin{Theorem}\label{Padm-smooth:thm}
  If $(\varepsilon,S)\in L^1(\cT^*)$, then $\bPadm$ defined by
  \bref{Eadm:def},\bref{padm:def} defines a smooth function on the
  Hilbert manifold $\cC(\varepsilon,S)$,
\[
  \bPadm \in C^\infty(\cC(\varepsilon,S),\bR^{3,1}).
\]
\end{Theorem}

\Proof We begin by proving an analogue of the $L^2$ bounds
\bref{Phi0-L2:bnd}, \bref{Phii-L2:bnd}.
\begin{Proposition}\label{Phi-L1bnd:thm}
  Suppose $g\in\cGpl$ for some $\lambda > 0$ and $\pi\in\cK$.  There
  is a constant $c=c(\lambda)$ such that
        \begin{eqnarray}  
            \nm{\Phi_0(g,\pi) - \cRo(g)}{L^1(\cM)}
              & \le & c\,\bigl(1    
                          +\nm{\grado g}{2,-3/2}^2 + \nm{\pi}{2,-3/2}^2  \nonumber \\ 
              &&\phantom{c\,\bigl(1}+ \nm{g-\go}{2,-1/2}\nm{\grado^2g}{2,-5/2} \bigr),
        \label{Phi0-L1:bnd} \\
                \nm{\Phi_{i}(g,\pi)-\cPo_{i}(\pi)}{L^1(\cM)}
                  & \le & c\,\left(\nm{g-\go}{2,-{1/2}} \right.
                                   \nm{\grado\pi}{2,-{5/2}}  \nonumber \\
             &&\phantom{c\,\bigl(}\left. + \nm{\grado g}{2,-3/2} 
                                           \nm{\pi}{2,-3/2}  \right).
        \label{Phii-L1:bnd}
        \end{eqnarray}
\end{Proposition}

\Proof
From \bref{Rg-go:eqn} we may express the scalar curvature in 
terms of $\cRo(g)$ by
\begin{eqnarray}
        R(g) & = & \cRo(g)/\rto +Q(g^{-1},\grado g) + g^{jk}\Ric(\go)_{jk}
                  \nonumber \\  
         &  & {} + \left(\left(g^{ik}-\go^{ik}\right)g^{jl}
                               + \go^{ik}\left(g^{jl}-\go^{jl}\right)\right)
                       \left(\grado^2_{ij}g_{kl} - \grado^2_{ik}g_{jl}\right) ,
        \label{Rg-Ro:eqn} 
\end{eqnarray}
The individual terms may be easily estimated as before, giving
\begin{equation}
        \nm{R(g)\rtg-\cRo(g)}{L^1(\cM)} \le c(\lambda)
            \left(1+\nm{g-\go}{2,-1/2} \nm{\grado^2g}{2,-5/2} 
                   + \nm{\grado g}{2,-3/2}^2 \right),
        \label{Rg-L1:bnd}
\end{equation}
from which \bref{Phi0-L1:bnd} follows, since
\begin{displaymath}
        \nm{\len{\pi}_g^2}{L^1(\cM)} \le c(\lambda) \nm{\pi}{2,-3/2}^2.
\end{displaymath}

From \bref{Phii-go:def} it follows that
\[
        \Phi_i(g,\pi) - \cPo_i(\pi) = \left(g_{ij} - \go{}_{ij}\right)
             \grado_k\pi^{jk} + g_{ij} A^j_{kl}\pi^{kl},
\]
which can be bounded  easily,
\begin{eqnarray}
\lefteqn{\nm{\Phi_i(g,\pi) - \cPo_i(\pi)}{L^1(\cM)} }\nonumber \\
        &\le &
          c\,\left(\nm{g-\go}{2,-1/2} \nm{\grado\pi}{2,-5/2} 
             +\nm{\grado g}{2,-3/2} \nm{\pi}{2,-3/2}\right),
        \label{Phii-Po:est}
\end{eqnarray}
as required.
\qed

Since $\bPadm(g,\pi)$ depends linearly on $(g,\pi)$, 
to complete the proof of Theorem \ref{Padm-smooth:thm} it will suffice to 
show that $\bPadm$ is bounded on $\cC(\varepsilon,S)$.
From \bref{Phi0-L1:bnd} we see that
\begin{eqnarray*}
        \nm{\cRo(g)}{L^1} & \le & \nm{\Phi_0(g,\pi)-\cRo(g)}{L^1}
               + \nm{\Phi_0(g,\pi)}{L^1}
                  \\
         & \le & c(g) \bigl(1+\nm{\grado g}{2,-3/2}^2 
              + \nm{\pi}{2,-3/2}^2  
              \\
     & &\phantom{c(g) \bigl(1}
              + \nm{g-\go}{2,-1/2}\nm{\grado^2g}{2,-5/2}\bigr)
              + \nm{\varepsilon}{L^1},
\end{eqnarray*}
and hence $\cRo(g)$ is integrable.  
Since $\grado\xhinf$ has compact support, it follows that the integrand of 
\bref{Eadm:def} is integrable and $\bPadm_0(g,\pi)$ is finite on $\cC(\varepsilon,S)$.
Similarly we estimate using \bref{Phii-L1:bnd}, assuming 
$\len{\xhinf^i}\le1$ for simplicity,
\begin{eqnarray*}
        \nm{\xhinf^i\cPo_i(\pi)}{L^1} & \le & 
                 \nm{\xhinf^i(\Phi_i(g,\pi)-\cPo_i(g))}{L^1}
               + \nm{\xhinf^i\Phi_i(g,\pi)}{L^1}
                  \\
         & \le & c \left(\nm{g-\go}{2,-1/2}\nm{\grado\pi}{2,-5/2}
              + \nm{\grado g}{2,-3/2} \nm{\pi}{2,-3/2}  \right)
              \\
         &&{}     + \nm{S}{L^1},
\end{eqnarray*}
whereupon the integrand of \bref{padm:def} is integrable and thus 
$\bPadm_i(g,\pi)$ is finite.  
\qed

We now show that the definitions \bref{Eadm:def}, \bref{padm:def}
adopted for $\bPadm$ agree with the formal definitions
\bref{E-formal:def}, \bref{p-formal:def}, when suitably interpreted,
under the general conditions of the mass existence Theorem
\ref{Padm-smooth:thm}, and that the value of $\bPadm(g,\pi)$ does not
depend on the choice of structure of infinity $\phi$ and its
associated background metric $\go = \phi^*(\delta)$
cf.~\cite[Theorem 4.2]{Bartnik86}.

The following two elementary lemmas will take care of the major technical 
details of the proof, and will be useful elsewhere.  The first lemma 
reviews the validity of integration by parts, and is valid under 
considerably more general circumstances than required here.

\begin{Lemma}\label{integration:lem}
        Suppose $\cM = \bigcup_{k\ge1}\cM_k$ is an exhaustion of a non-compact, 
        $n$-dimensional manifold $\cM$ by 
        compact subsets with smooth boundaries $\partial\cM_k$, and suppose 
        $\beta\in W^{1,2}_{loc}(\Lambda^{n-1}T^*\cM)$ satisfies 
        $d\beta\in L^1(\Lambda^nT^*\cM)$.  Then
        \begin{itemize}
                \item[(i)]  $$\oint_{\partial\cM_k} \beta\quad
                               \hbox{  exists for $k\ge1$};$$
                \item[(ii)]  $$\oint_{\partial\cM_\infty}\beta := 
                \lim_{k\rightarrow\infty}\oint_{\partial\cM_k}\beta \hbox{  exists.}$$
        \end{itemize}
\end{Lemma}

\Proof
   Since $\partial\cM_k$ is smooth, the trace theorem \cite{Stein70,Taylor81}  
   shows that $\beta\in W^{1/2,2}(\partial\cM_k)\subset L^2(\partial\cM_k) 
   \subset L^1(\partial\cM_k)$, where the fractional Sobolev space is 
   defined using the Fourier transform in the usual manner.  
   This shows that the 
   finite boundary integrals are well-defined.  The definition of weak 
   derivative allows us to apply Stokes' theorem to $d\beta$ over any 
   compact region; in particular, for $1\le q\le p$ we have
   $$ 
      \oint_{\partial\cM_p} \beta - \oint_{\partial\cM_q} \beta  =
      \int_{\cM_p\backslash\cM_q} d\beta.
   $$
   Since $d\beta\in L^1(\Lambda^nT^*\cM)$, the right-hand side is $o(1)$ as
   $q = \min(p,q) \rightarrow \infty$ and hence 
   $\{\oint_{\partial\cM_k}\beta\}^\infty_{k=1}$ is a Cauchy sequence and 
   convergent as claimed.
\qed

Likewise, the second lemma is valid with more general values for   
the indices, but this will not be needed here.
\begin{Lemma} \label{u1SR:lem}
        Suppose $E_{R_0}\subset\bR^3$, $R_0\ge1$ and 
        $u\in W^{1,2}_{-3/2}(E_{R_0})$.
        Then $u\in L^4(S_R)$ for every $R\ge R_0$, and
        there is a constant $c$, independent of $R$, such that 
        \begin{equation}
                \oint_{S_R} \len{u}\,dS \le c R^{1/2}\nm{u}{1,2,-3/2;A_R},
                \label{u1SR:est}
        \end{equation}
        (where the notation indicates the norm over the annulur domain $A_R$);
        hence
        \begin{equation}
                \nm{u}{1;S_R} = o(R^{1/2})\quad\hbox{as $R\rightarrow\infty$}.
                \label{u1SR:dcy}
        \end{equation}
  \end{Lemma}  
\Proof 
  As in \cite{Bartnik86} we define $u_R(x) = u(Rx)$, and recall the
  uniform comparison
  \begin{displaymath}
    \nm{u_R}{k,p;A_1} \approx R^\delta \nm{u}{k,p,\delta;A_R},\quad 
    \hbox{for any $R\ge R_0$.}
  \end{displaymath}
  Since $u_R\in W^{1,2}(A_1)$, the trace theorem 
  again implies 
  $u_R\in W^{1/2,2}(S_1)$, and 
  \begin{equation}
        \nm{u_R}{1/2,2;S_1}  \le  \nm{u_R}{1,2;A_1}.
        \label{tracethm:est}
  \end{equation}
  It readily follows that
  \begin{displaymath}
        \nm{u_R}{1;S_1} \le c \nm{u_R}{4;S_1} \le c \nm{u_R}{1/2,2;S_1}
        \le \nm{u_R}{1,2;A_1}   
  \end{displaymath}  
  and thus 
  \begin{displaymath}
        \nm{u}{1;S_R} \le c R^2  \nm{u_R}{1,2;A_1}      \le c R^{1/2} 
        \nm{u}{1,2,-3/2;A_R}.
  \end{displaymath}
  In fact, using the Sobolev inequality in $W^{1/2,2}(S_1)$ gives
  $$
    \nm{u}{4;S_R} \le c R^{-1} \nm{u}{1,2,-3/2;A_R};
  $$
  a stronger inequality  which we will not need here.
  The conclusion \bref{u1SR:dcy} follows as in \cite{Bartnik86}, since 
  $u\in W^{k,p}_{\delta}(E_{R_0})$ implies both
  $\nm{u}{k,p,\delta;A_R} = o(1)$ and $\nm{u}{k,p;A_R}=o(R^{\delta})$
  as $R\rightarrow \infty$.
\qed

It follows easily that the formal asymptotic definition of 
$(E,p)$ agrees with the integral definition of $\bPadm$.  This 
generalises and extends Proposition 4.1 of \cite{Bartnik86}.

\begin{Proposition}\label{ADMdef-ok:pro}
  Suppose $(g,\pi)\in\Phi^{-1}(L^1(\cT^*\otimes\Lambda^3))$.  Then
  $(E,p)$ from \bref{E-formal:def}, \bref{p-formal:def} are defined,
  in the sense of Lemma \ref{integration:lem}, and satisfy $(E,p) =
  \bPadm$.
\end{Proposition} 

\Proof
   After noting that the integrals of \bref{Eadm:def}, \bref{padm:def} 
   may be written as exact divergences, respectively of
   \begin{eqnarray}\label{P-diverge:eqn}
         &  & \grado^i \left(\xhinf^0\left(\grado^j g_{ij} 
                            - \grado_i\tro g\right)\right)
              \rtg,  \nonumber \\
         &  & 2 \grado_i \left(\xhinf^k \go{}_{jk}\pi^{ij}\right),
   \end{eqnarray}
   which both satisfy the integrability condition of Lemma 
   \ref{integration:lem}, by Proposition \ref{Phi-L1bnd:thm} and the 
   hypothesis $\Phi(g,\pi) \in L^1(\cT^*\otimes\Lambda^3)$, we see that 
   $(E,p)$ is well-defined.  The equality of the two definitions is now a 
   tautology.
\qed

\begin{Corollary}\label{P-gendef:cor}
  The definition \bref{Eadm:def}, \bref{padm:def} of
  $\xi^\alpha_\infty\bPadm_\alpha(g,\pi)$ remains valid (and
  unchanged) if the constant translation at infinity $\xi_\infty$ is
  replaced by any asymptotic translation
  $\xi\in\xinf+W^{2,2}_{-1/2}(\cT)$.
\end{Corollary}
\Proof
    The difference between the two definitions of $\bPadm$ (using 
    $\xhinf, \xi$ respectively) is a sum of divergences of the form 
    \bref{P-diverge:eqn}, with $\xhinf$ replaced by $\xi-\xhinf \in 
    W^{2,2}_{-1/2}(\cT)$.  The weighted Sobolev inequality implies 
    $\xi-\xhinf$ is H\"older continuous and decays as $o(R^{-1/2})$, so 
    by Lemma \ref{u1SR:lem}, the boundary integral of \bref{P-diverge:eqn}
    is defined and decays as $o(R^{-1/2}) o(R^{1/2}) = o(1)$.
\qed

The proof that the value of $\bPadm$ is independent of the choice of
structure of infinity $\phi$ follows \cite[Theorem 4.2]{Bartnik86}.

\begin{Theorem}\label{ADM-uniqueness:thm}
  Suppose $\phi:\cM\backslash\cM_0\rightarrow\bR^3$,
  $\psi:\cM\backslash\cM_1\rightarrow\bR^3$ are two structures of
  infinity such that $(g,\pi)\in\cF(\phi)\cap\cF(\psi)$, where the
  notation indicates the phase space (and weighted Sobolev spaces)
  defined with respect to the indicated structure of infinity.  Then
  $\cF(\phi) = \cF(\psi)$, the underlying Hilbert Sobolev spaces have
  comparable norms, and $\bPadm(g,\pi;\phi) = \bPadm(g,\pi;\psi)$.
\end{Theorem}

This justifies the notation used elsewhere in this paper, 
where we do not indicate the choice of structure of infinity.

\Proof
   If  $g\in\cGp$ then by the Sobolev inequality, $\phi_*g-\delta\in 
   W^{1,6}(E_R)$, and $\phi,\psi$ satisfy the conditions of 
   \cite[Section 3]{Bartnik86}.  Hence the transition function 
   $\psi\circ\phi^{-1}:E_{R_2}\rightarrow\bR^3$ for some $R_2\ge1$, 
   after possibly moving $\psi$ by a rigid motion of $\bR^3$,
   satisfies $\psi\circ\phi^{-1}-\hbox{Id} \in W^{2,6}_{1/2}(E_{R_2})$.
   A trivial modification shows $\psi\circ\phi^{-1}-Id \in 
   W^{3,2}_{1/2}(E_{R_2})$, whereupon the background metrics satisfy
   $\phi^*\delta - \psi^*\delta \in W^{2,2}_{-1/2}(\cS(\cM_0\cap\cM_1))$
   and it follows that the spaces $\cGp, \cK$ are in fact 
   independent of the choice of structure of infinity $\phi$.
   
   To show invariance of the ADM energy-momentum, let $\gt$ be a 
   background metric for $\psi$, so $\gt = \psi^*\delta$ in 
   $\cM\backslash\cM_1$, and let $y(x) = \psi\circ\phi^{-1}(x)$, $x\in 
   E_{R_2}$ be the coordinate transition function.  Let $\gradt$ and
   $\tilde{\bPadm}$, respectively, be the 
   connection and total ADM energy-momentum operators of $\gt$.   
   By Corollary \ref{P-gendef:cor} and the uniqueness of $W^{2,2}_{-1/2}$, 
   we may use the same vector field $\xi\in\xinf+W^{2,2}_{-1/2}(\cT)$ 
   to define both $\bPadm, \tilde{\bPadm}$.
   
   The divergence expression \bref{P-diverge:eqn} for the integrand for 
   $\xinf^\alpha\bPadm_\alpha$ 
   may be written in arbitrary coordinates in the form
   $$
   \del_p\left(\xi^0\go^{ip}\go^{jk}\left(\grado_kg_{ij} - 
   \grado_ig_{jk}\right)\rto\right) 
   + 2 \del_i\left(\xi^k\go{}_{jk}\pi^{iij}\right),
   $$
   with a similar expression being valid for 
   $\xinf^\alpha\tilde{\bPadm}_\alpha$.
   Since 
   $$
      \grado_ig_{jk}-\gradt_ig_{jk} = \tilde{A}^p_{ij}g_{pk} -
              \tilde{A}^p_{ik}g_{jp},
   $$
   where $\tilde{A}^p_{ij}=\tilde{\Gamma}^p_{ij} - \Gammao^p_{ij}$,
   after a certain amount of calculation we find that the difference 
   between the energy-momentum integrands may be written in the form
   \begin{eqnarray*}
         &  & \del\left(\phantom{{}^0}
                   \xi(\go-\gt)(\pi+\grado g) + \xi(g-\go)\grado \gt\right)  \\
         & + &  \del_j \left(\xi^0 \left(\grado^j \tro\gt - 
         \go^{jk}\grado^l\gt_{kl}\right)\rto\right).
   \end{eqnarray*}
   The precise form of the first term is of no account, since by 
   the argument of Corollary \ref{P-gendef:cor} and the decay conditions 
   on $g,\gt,\xi$, the first term integrates to zero.  Integrating, we arrive 
   at the relation
   \begin{equation}
     \xi^\alpha\bPadm_\alpha(g,\pi) - \xi^\alpha\tilde{\bPadm}_\alpha(g,\pi)
              = \xi^\alpha\bPadm_\alpha(\gt,0),
        \label{P-change:eqn}
   \end{equation}
   and it remains to show that $(\gt,0)$ has vanishing energy-momentum
   (notice that the fact  $\gt = \psi^*(\delta)$ 
   has not yet been used, so \bref{P-change:eqn} is valid more generally).
   
   Working in the $\go$ rectangular coordinates $x^i$, in which 
   $\go{}_{ij}=\delta_{ij}$, the metric $\gt$ is given in terms of the 
   transition functions $y(x)$ by $\gt_{ij} = \del_i y^p \del_j y^p$, 
   where $\del_i = \del/\del x^i$.  Since $\go$ is explicitly flat in the 
   coordinates $x^i$, $\xi^\alpha\bPadm_\alpha(\gt,0) = 
   \xi^0\bPadm_0(\gt,0)$ is the integral of the $\bR^3$-divergence of 
   $$
         \xi^0 (\del_i\gt_{ij} - \del_j\gt_{ii}) 
         = \xi^0(\del^2_{ii}y^p \del_jy^p -\del_iy^p \del^2_{ij}y^p).
   $$
   After a rotation, we may assume $\del y^p/\del x^i - \delta^p_i \in 
   W^{2,2}_{-1/2}$ in the exterior region, and therefore by the argument 
   of Corollary \ref{P-gendef:cor} again, the above expression may be 
   reduced to
   $$
   \del^2_{ii}y^j - \del^2_{ij}y^i.
   $$
   Expressing this explicitly as a 2-form gives
   \begin{eqnarray*}
        \left(\del^2_{ii}y^j - \del^2_{ij}y^i\right)*\!dx^j & = & 
           \del_i\left(\del_iy^j - \del_jy^i\right) *\!dx^j  \\
         & = & -d\left(\epsilon_{ijk}\del_i y^j\,dx^k\right),
   \end{eqnarray*}
   which is a closed 2-form and therefore does not contribute to any 
   boundary integral.  It follows that $\bPadm(\gt,0) = 0$.
\qed

\section{Hamiltonians}\label{hamiltonian:sec}
%
%
The formal variational structure of the Einstein equations is well-known 
and due originally to Hilbert and Einstein 
\cite{Einstein16,Hilbert15}: 
the Euler-Lagrange equations of the Lagrangian functional
\begin{equation}
        \cLEH(g^{(4)}) := \int_{V} R(g^{(4)}) \sqrt{g^{(4)}} \, d^4x,
        \label{EHlagrangian:def}
\end{equation}
are obtained in the usual manner, by making a compactly supported
variation of the spacetime metric $g^{(4)}$ and once integrating by
parts, and are just the (vacuum) Einstein equations.  In this respect
the Einstein equations are similar to the equations of motion of most
other Lagrangian field theories, such as the classical wave equation.
However, it differs in that although the resulting equations are
second order in the metric, the Lagrangian contains explicit second
derivatives.  As is well known, the Gauss-Bonnet formula shows that
the Einstein-Hilbert integrand can be written in a local coordinate
system in the form
\[
R(g^{(4)}) \sqrt{g^{(4)}} \, d^4x =d(A_1(g^{(4)},\partial g^{(4)})) +
A_2(g^{(4)},\partial g^{(4)}),
\]
where $A_1$ is linear and $A_2$ is quadratic in $\partial g^{(4)}$,
and thus the Euler-Lagrange equations are determined by $A_2$ since
compactly supported variations of the divergence terms $dA_1$ will not
contribute to the equations.  However, $A_2$ is neither unique nor a
geometrically invariant quantity and we therefore have the curious
situation of a non-unique, non-geometric (coordinate dependent),
integrand giving rise to a geometric (tensorial) Euler-Lagrange
equation.

The Hamiltonian interpretation of the Einstein-Hilbert Lagrangian was 
provided by Arnowitt, Deser and Misner \cite{ADM61}, who decomposed
$\cLEH$ by imposing a $3+1$ splitting of the spacetime $V$ and after 
an integration by parts in the time direction and dropping the resulting 
boundary integral, arrived at the ADM form of the Lagrangian
\begin{equation}
   \cLEH \simeq 
   \int_{V}\left( \pi\sbullet\del_tg - \xi^\alpha\Phi_\alpha(g,\pi)\right)
        \label{ADMlagrangian:def}
\end{equation}
where $\xi=(N,X^i)$ is the (unspecified) lapse and shift of the $3+1$ 
decomposition.  This decomposition, incidentally, is the origin of the form
\bref{pi:def} for the conjugate momentum $\pi$.  Now introducing the ADM 
Hamiltonian, 
\begin{equation}
        \cHadm(g,\pi;\xi) = - \int_{\cM} \xi^\alpha \Phi_\alpha(g,\pi),
        \label{Hadm:def}
\end{equation}
the Einstein-Hilbert variational computation, with compactly supported 
variations, may be re-expressed as Hamilton's equations of motion for 
$\cHadm$,
\begin{equation}
        {d\over dt}\left( \begin{array}{c} g  \\ \pi  \end{array} \right)
        = - J\ D\Phi(g,\pi)^*(\xi)
        \label{ADM:eqn}
\end{equation}
where 
$J=\left(
\begin{array}{cc}
        0 & 1  \\
        -1 & 0
\end{array}
\right):T\cF \rightarrow T\cF$ is the implied symplectic form, 
  $J\left(\begin{array}{c}-p\\ h\end{array}\right) = 
    \left(\begin{array}{c}h\\ p\end{array}\right)$, 
in the $(g,\pi)$ coordinates on $\cF$ \cite{FischerMarsden79,ChoquetYork79}.

We note parenthetically that this $3+1$ reduction involves two
geometric (gauge) choices; that of a timeflow vector field (reducing
to $\xi$ on the hypersurface) and a choice of spacelike hypersurface.
Rather remarkably, it turns out that the spacelike integrand may be
considered as the restriction to the hypersurface of a $3$-form
defined globally on the spacetime, and depending only on the spacetime
metric and the choice of timeflow vector field --- see
\cite{Nester83} for the computations involved.

In this section we are instead concerned with formulating the above
equations in the context of the phase space $\cF$.  The aim is to
construct Hamiltonian functionals which, together with apropriately
chosen decay and boundary conditions for the lapse-shift $\xi$, lead
to the evolution equations \bref{ADM:eqn}.  The existence and
uniqueness for the Einstein evolution equations is a separate and
rather difficult question in analysis which will not be considered
here --- in particular, it does not seem possible to deduce this on
general grounds from the Hamiltonian structure on the phase space.

Since one of the primary difficulties is the control of boundary
terms, we record the complete form of the boundary terms arising from
the integration by parts relating the variational derivative
$D\Phi(g,\pi)$ to the adjoint operator $D\Phi(g,\pi)^*$.  This follows
directly from the expressions (\ref{DPhi0*}--\ref{DPhi*}).
\begin{eqnarray} 
\lefteqn{\xi^\alpha D\Phi_\alpha(g,\pi)(h,p)
    - D\Phi(g,\pi)^*\xi\cdot (h,p) }\qquad\qquad\nonumber \\ 
    &= & 
    \grad^i\left\{\xi^0\left(\grad^jh_{ij}-\grad_i\trg h\right)
    - \left(h_{ij}\grad^j\xi^0-\trg h \grad_i\xi^0\right)\right\}\rtg
    \nonumber \\  
    &  & {}+\grad_i\left\{2\xi_jp^{ij}+2\xi^j\pi^{ik}h_{jk}-
                           \xi^i\pi^{jk}h_{jk}\right\} 
    \label{DPhiBndry:eqn} 
\end{eqnarray} 
In the following we assume that $\del\cM$ is empty.

\begin{Theorem}\label{Hadm:thm}
  The ADM Hamiltonian \bref{Hadm:def} with lapse-shift $\xi \in\cL$
  defines a smooth map of Hilbert manifolds
  \begin{displaymath}
                \cHadm : \cF\times\cL \rightarrow \bR.
  \end{displaymath}
  If $\xi\in W^{2,2}_{-1/2}(\cT)$, then for all $(h,p)\in
  T_{(g,\pi)}\cF$,
  \begin{equation}
                D_{(g,\pi)}\cHadm(g,\pi;\xi)(h,p) 
                  = -\int_{\cM} (h,p) \cdot D\Phi(g,\pi)^*(\xi).
                \label{DHadm:eqn}
  \end{equation}
\end{Theorem}
\Proof
  H\"older's inequality and the decay condition $\xi\in \cL = 
  L^2_{-1/2}(\cT)$ shows that $\cHadm$ is defined and bounded on 
  $\cF\times\cL$, hence the linearity in $\xi$ implies smoothness with 
  respect to $\xi$. Likewise, smoothness with respect to $(g,\pi)$ follows 
  from the smoothness of the map $(g,\pi)\mapsto\Phi(g,\pi)$. 
  
  To show \bref{DHadm:eqn} we must control the boundary terms in
  \bref{DPhiBndry:eqn}.  For this we use the trace theorem, in the
  form of Lemma \ref{u1SR:lem}.  The individual components of the
  boundary term
  \begin{eqnarray}
        B^i &=& \xi^0\left(\grad^jh_{ij}-\grad_i\trg h\right)\rtg
            - \left(h_{ij}\grad^j\xi^0-\trg h \grad_i\xi^0\right)\rtg , 
       \nonumber      \\
             &&{}+ \xi_jp^{ij}+\xi^j\pi^{ik}h_{jk}-  \half\xi^i\pi^{jk}h_{jk}  
        \label{DPhi-bndryterms:def}
  \end{eqnarray}
  have well-defined traces on the spheres $S_R$ (and on any other
  smooth hypersurface in $\cM$), thus integration of the adjoint
  operator formula \bref{DPhiBndry:eqn} yields the expected boundary
  integrals.  We may now estimate the boundary contribution over $S_R$
  in the limit as $R\rightarrow\infty$,
  \begin{eqnarray*}
        \oint_{S_R} \len{B}\,dS& \le & \phantom{{}+} \nm{\xi}{\infty;S_R} 
      \left(\nm{\grad h}{1;S_R} + \nm{p}{1;S_R}\right)  \\
         &  & {}+ \nm{h}{\infty;S_R}
                \left(      \nm{\grad \xi}{1;S_R}  + 
                        \nm{\xi}{\infty;S_R}\nm{\pi}{1;S_R}\right)
      \\
         & \le & o(1)\left(
            \nm{\grad h}{1,2,-3/2;A_R} 
            + \nm{\grad\xi}{1,2,-3/2;A_R}  \right.  \\
         &&\phantom{o(1)\bigl(} \left.
            + \nm{\pi}{1,2,-3/2;A_R}
            + \nm{p}{1,2,-3/2;A_R} \right),
  \end{eqnarray*}
  which shows the boundary integral is $o(1)$ as $R\rightarrow\infty$.
  Integrating \bref{DPhiBndry:eqn} over $\cM_R :=
  \{x\in\cM:\sigma(x)<R\}$ and letting $R\rightarrow\infty$
  establishes \bref{DHadm:eqn} and completes the proof of Theorem
  \ref{Hadm:thm}.  Note that the use of the spherical exhaustion
  $\cM_R, R\ge R_0$, is merely a convenience; since the integrands
  $\xi D\Phi(h,p)$, $(h,p)\cdot D\Phi^*\xi$ are integrable, the
  improper integrals in \bref{DHadm:eqn} are independent of the choice
  of exhaustion used to define them, and the boundary integrals
  evaluated with any other smooth exhaustion of $\cM$ will also vanish
  in the limit.  
\qed

We emphasise that an identity of the form \bref{DHadm:eqn} is
necessary if the Hamiltonian is to generate the correct equations of
motion.  The restriction above to lapse-shift $\xi$ decaying at
infinity is essential, both in defining $\cHadm$ (since $\Phi(g,\pi)$
is not integrable for generic $(g,\pi)\in\cF$) and in ensuring that
asymptotic boundary terms are absent in \bref{DHadm:eqn}.  However, we
would like to be able to choose $\xi$ asymptotic to a translation at
infinity in the evolution equations and retain the validity of
\bref{DHadm:eqn}; this necessitates a modification of the Hamiltonian
functional $\cHadm$, as suggested in \cite{ReggeTeitelboim74}.

The underlying principle here is that adding a divergence to the
Hamiltonian (or the Einstein-Hilbert Lagrangian) will not change the
formal equations of motion, but such a term will affect the phase
space (domain of definition) of the Hamiltonian and the resulting
equations of motion.  In particular, to extend the definition of the
ADM Hamiltonian to permit lapse-shift asymptotic to a (non-zero)
translation at infinity, we should add a divergence which cancels the
dominant contribution from the asymptotic translation --- from
\bref{Eadm:def}, \bref{padm:def} we recognise that the ADM energy
$\xinf^\alpha\bPadm_\alpha$ is an appropriate choice.  Thus we arrive
at the Regge-Teitelboim Hamiltonian \cite{ReggeTeitelboim74}
\begin{equation}
        \cHrt(g,\pi;\xi) = 16\pi \xinf^\alpha\bPadm_\alpha(g,\pi) - 
          \int_\cM \xi^\alpha\Phi_\alpha(g,\pi),
        \label{Hrttemp:def}
\end{equation}
where $\xi\in \xinf + \cL$.  This expression is well-defined on $\cC$,
where it has the value $\xinf^\alpha\bPadm_\alpha(g,\pi)$, and more
generally on $\Phi^{-1}(L^1(\cT^*))$, but for general $(g,\pi)\in\cF$
the individual terms are not defined, and thus \bref{Hrttemp:def} does
not provide a definition valid on all $\cF$.  We circumvent this
problem by inserting the definition of $\bPadm$ and rearranging terms
--- thus for general $(g,\pi)\in\cF$ and $\xi\in\xinf+\cL$ we define
the regularised Hamiltonian $\cH(g,\pi,\xi)$ by
\begin{eqnarray}
\lefteqn{\cH(g,\pi;\xi)  =  }\qquad\qquad\nonumber \\*
        &&\phantom{{}+}
                \int_\cM \left(\xhinf^0-\xi^0\right) \Phi_0(g,\pi)
              + \int_\cM \left(\xhinf^i-\xi^i\right) \Phi_i(g,\pi)
        \nonumber \\*
          &&{}+ \int_\cM \xhinf^0\left(\cRo(g)-\Phi_0(g,\pi)\right)
              + \int_\cM 
              \grado^i\xhinf^0\left(\grado^jg_{ij}-\grado_i\tro g\right)\rto 
        \nonumber \\*
          &&{}+ \int_\cM \xhinf^i\left(\cPo_i(\pi)-\Phi_i(g,\pi)\right)
              + \int_\cM 2\pi^{ij}\grado_i\xhinf{}_j ,
        \label{H:def}
\end{eqnarray}
where $\xi\in \xinf + \cL$ and $\xhinf$ is constant at infinity with
value $\xinf$.  For $(g,\pi)\in\Phi^{-1}(L^1(\cT^*))$ this agrees with
\bref{Hrttemp:def}.  As in Section \ref{ADM:sec}, the sum of the
integrals in \bref{H:def} is independent of the particular choice of
constant at infinity vector field $\xhinf$ representing the
translation $\xinf$, although the pointwise values of the integrands
are not invariant.  We emphasise that for generic $(g,\pi)\in\cF$,
$\cH$ does not have a simple geometric interpretation such as
\bref{Hrttemp:def}.  Nevertheless, it does have some useful
properties.

\begin{Theorem}\label{HRT:thm}
        The functional $\cH(g,\pi;\xi)$ defined by \bref{H:def} is bounded 
        on $\cF\times(\bR^{3,1}+\cL)$ and smooth with respect to the Hilbert 
        structure on this space.
        If $\xi\in\xinf+W^{2,2}_{-1/2}(\cT)$ then for all $(g,\pi)\in\cF$ and 
        $(h,p)\in T_{(g,\pi)}\cF$  we have
        \begin{equation}
                D_{(g,\pi)}\cH(g,\pi;\xi)(h,p) = - \int_\cM (h,p)\cdot 
                D\Phi(g,\pi)^*\xi.
                \label{H-ibyp:eqn}
        \end{equation}
\end{Theorem}
\Proof
   As before, for smoothness it suffices to show that $\cH$ is bounded on 
   $\cF\times(\bR^{3,1}+\cL)$.  Since $\nm{\xi-\xhinf}{2,-1/2}\le C$ for 
   $\xi\in\xinf+\cL$, the first two integrals of \bref{H:def} may be 
   estimated by
   \begin{displaymath}
          \left|\int_\cM 
               \left(\xhinf^\alpha-\xi^\alpha\right)\Phi_\alpha(g,\pi)\right|
          \le \nm{\xi-\xhinf}{2,-1/2} \nm{\Phi(g,\pi)}{2,-5/2},
   \end{displaymath}
   which is bounded, by Theorem \ref{Phi-L2bnd:thm}.  
   The fourth and sixth integrals are 
   bounded because $\grado\xhinf$ has compact support, and the third and
   fifth integrals are bounded, since Proposition \ref{Phi-L1bnd:thm} shows that 
   $\cRo(g)-\Phi_0(g,\pi)$ and $\cPo_i(\pi)-\Phi_i(g,\pi)$ are both 
   integrable ($L^1(\cM)$). 
   Hence $\cH$ is bounded and therefore smooth, 
   by the same arguments as used in Proposition \ref{Phi-L2bnd:thm}.  
   To show \bref{H-ibyp:eqn} we must separately consider the 
   variational derivatives of the individual terms of \bref{H:def}.
   Since $\xi-\xhinf \in W^{2,2}_{-1/2}(\cT)$, Theorem \ref{Hadm:thm} 
   may be applied to the variation of the first two integrals, 
   which may then be rewritten as
   \begin{equation}
          \int_\cM (h,p)\cdot D\Phi(g,\pi)^*(\xhinf-\xi).
          \label{dh1:exp}
   \end{equation}
   The variational derivative of the third and fourth terms of   
   \bref{H:def} may be rearranged using \bref{Ro:def}, \bref{DPhi0:def}, 
   \bref{DPhiBndry:eqn} to give
   \begin{eqnarray*}
        \lefteqn{ \int_\cM \left\{ \grado^i\left( 
              \xhinf^0\left(\grado^jh_{ij}-\grado_i\tro h\right)\right)\rto
              - \xhinf^0 D\Phi_0(g,\pi)(h,p) \right\}
              }\qquad
    \\
         & = & \int_\cM\left\{ 
              \grado^i\left(\xhinf^0\left(\grado^jh_{ij}
                               -\grado_i\tro h\right)\right)\rto
             -\grad^i \left(\xhinf^0\left(\grad^jh_{ij}
                               -\grad_i\trg h\right)\right)\rtg     \right.
    \\
         &  & \phantom{\int_\cM\bigl\{ } \left.
              + \grad^i\left(\grad^j\xhinf^0 h_{ij} 
                              - \grad_i \xhinf^0 \trg h\right)\rtg
              - (h,p)\cdot D\Phi_0(g,\pi)^*(\xhinf^0)   \right\}.
   \end{eqnarray*}
   The dominant terms of the first two divergences in this expression cancel, 
   and the remaining parts of the 
   boundary term may therefore be written symbolically as
   $\xhinf(g-\go)(\grado h + h\grado g)$.  Now 
   $\len{\xhinf}= O(1)$ and $g-\go = o(R^{-1/2})$, and
   Lemma \ref{u1SR:lem} serves to show that the remaining terms have 
   well-defined traces, hence the boundary integral is $o(1)$ as 
   $R\rightarrow\infty$.  Consequently the variation of the third and 
   fourth terms of \bref{H:def} is just 
   \begin{displaymath}
            - \int_\cM  (h,p)\cdot D\Phi_0(g,\pi)^*(\xhinf^0).
   \end{displaymath}
   The argument controlling the variational derivative of the final 
   two terms of \bref{H:def} is very similar, and results in the 
   expression 
   \begin{displaymath}
            - \int_\cM  (h,p)\cdot D\Phi_i(g,\pi)^*(\xhinf^i),
   \end{displaymath}
   from which the final identity \bref{H-ibyp:eqn} follows.
\qed
  
\section{Critical points of the ADM mass}
\label{sec:critical}
%
%

The results of the previous section, particular Theorem \ref{HRT:thm},
have an elegant interpretation in terms of critical points of the ADM
mass.  The fundamental observation is that stationary metrics are
critical points of the ADM energy functional on the constraint
manifold; and an argument implying the converse was suggested in
\cite{BrillDeserFadeev68}.  
In this section we show
that the phase space $\cF$ and the regularised Hamiltonian functional
$\cH$ allow a rigorous presentation of the previously heuristic
arguments relating stationary metrics and criticality properties of
the ADM mass.  The main result establishes the equivalence between
critical points of the total energy and generalised Killing vectors.

\begin{Theorem}\label{Ecrit:thm}
  Suppose $(g,\pi)\in\cF$ satisfies $\Phi(g,\pi) =
  (\varepsilon,S_i)\in L^1(\cT^*\otimes\Lambda^3)$, let $\xinf \in
  \bR^{3,1}$ be a fixed future timelike vector and define the energy
  functional $E\in C^{\infty}(\cC(\varepsilon,S_i))$ by
  \begin{equation}
           E(g,\pi) = \xinf^\alpha \bPadm_\alpha(g,\pi),
                     \quad \forall\ (g,\pi)\in\cC(\varepsilon,S_i).
  \label{Energy:def}
  \end{equation}
  Then the following two statements are equivalent:
    \begin{itemize}
      \item[(i)]  For all $(h,p)\in T_{(g,\pi)}\cC(\varepsilon,S_i)$ we have 
             \[  DE(g,\pi)(h,p) = 0 ;  
             \]
      \item[(ii)]  
           There is $\xi\in\xi_\infty+W^{2,2}_{-1/2}(\cT)$ satisfying
                \[ D\Phi(g,\pi)^*\xi = 0 . \] 
        \end{itemize}
\end{Theorem}

If the energy-momentum covector $\bPadm$ is timelike or null then the ADM 
(total) mass can be defined,
\[
\madm = \sqrt{- \bPadm^\alpha\bPadm_\alpha},
\]
and in many applications, such as the quasi-local mass definition of
\cite{Bartnik89}, it is more natural to use $\madm$ rather than the
energy $E(g,\pi)$ with respect to the direction $\xi_\infty$.  The
following corollary shows how Theorem \ref{Ecrit:thm} can be used to
relate critical points of $\madm$ to stationary metrics.  The
hypothesis that $\bPadm$ be timelike follows from the extension in
\cite{BartnikChrusciel04} of the spinorial proof \cite{Witten81} of
the Positive Mass Theorem \cite{SchoenYau79,SchoenYau81,Witten81} to
the decay and regularity condition $(g,\pi)\in\cC(\varepsilon,S_i)$,
assuming that the local energy-momentum density $(\varepsilon,S_i)$
satisfies the Dominant Energy Condition
\[ 
   \xi^0 \varepsilon + \xi^i S_i \ge 0,\quad 
   \hbox{for all future timelike vector fields $\xi\in C^{\infty}_c(\cT)$.}
\]
Similarly, it is well-known that if $\xi$ is a Killing vector,
timelike near infinity, then $\bPadm^\alpha$ and $\xinf^\alpha$ are
proportional \cite{BeigChrusciel96a}. 

\begin{Corollary}\label{mcrit:cor}
   Suppose $(g,\pi)\in\cF$, $\Phi(g,\pi)=(\varepsilon,S_i)\in L^1(\cT^*)$ and 
   $\bPadm=\bPadm(g,\pi)$ is a future timelike vector.   
   If $D\madm(g,\pi)(h,p) = 0$ for all $(h,p)\in T\cC(\varepsilon,S_i)$,
   then $(g,\pi)$ is a generalised stationary initial data set, with 
   generalised Killing vector $\xi$ such that $\xi_{\infty}^\alpha$ is 
   proportional to $\bPadm^\alpha=\eta^{\alpha\beta}\bPadm_\beta(g,\pi)$.
   Conversely, if $(g,\pi)$ is a generalised stationary initial data set, 
   with generalised Killing vector $\xi$ such that $\xi_{\infty}^\alpha$ is 
   proportional to $\bPadm^\alpha$, then $D\madm(g,\pi)(h,p) = 0$ for 
   all $(h,p)\in T\cC(\varepsilon,S_i)$.
\end{Corollary}

\Proof
  If $\bPadm_\alpha$ is a timelike vector, then we may choose 
  $\madm\xi_\infty^\alpha = -\eta^{\alpha\beta}\bPadm_\beta$, thereby 
  normalising $\xinf$ to be a future unit timelike vector. Defining 
  $E=\xinf^\alpha\bPadm_\alpha$, we have 
  $D\madm = \xinf^\alpha D\bPadm_\alpha = DE$, and $\madm$ is 
  critical on $\cC(\varepsilon,S_i)$ exactly when $E$ is critical also.  
  Thus if $(g,\pi)$ is a critical point for $\madm$ on $\cC(\varepsilon,S_i)$
  then Theorem \ref{Ecrit:thm} shows that $(g,\pi)$ admits a generalised 
  Killing vector $\xi\in\xinf+W^{2,2}_{-1/2}$, with $\xinf$ proportional 
  to $(\bPadm^\alpha)$.
  
  Conversely, if $(g,\pi)$ admits a generalised Killing vector $\xi$ with 
  $\xinf$ proportional to $(\bPadm^\alpha)$, then defining 
  $E(g',\pi') = \xinf^\alpha\bPadm_\alpha(g',\pi')$ with $\xi$ 
  normalised so $\xinf$ is a unit timelike vector, it follows that 
  $DE(g,\pi)=0$ on $\cC(\varepsilon,S_i)$; since $D\madm=DE$, we then have 
  $D\madm(g,\pi)=0$ on $\cC(\varepsilon,S_i)$.
\qed

The proof of Theorem \ref{Ecrit:thm} is based on a generalisation of the 
classical method of Lagrange multipliers to Banach spaces, which we now 
recall. I am indebted to John Hutchinson for the following elegant proof. 

\begin{Theorem}\label{LagrangeMult:thm}
   Suppose $K:B_1\to B_2$ is a $C^1$ map between Banach spaces, such that
   $DK(u):B_1\to B_2$ is surjective and splits (ie.~$DK(u)$ has closed
   kernel, with closed complementary subspace), for every $u\in K^{-1}(0)$,
   and suppose $f\in C^1(B_1)$.  Let $u\in K^{-1}(0)$ be given, then the
   following are equivalent: \begin{itemize} \item[(i)] For all $v\in \ker
   DK(u)$ we have \[ Df(u)v = 0 ; \] \item[(ii)] There is $\lambda\in
   B_2^*$ such that for all $v\in B_1$, \[Df(u)v =
   \langle\lambda,DK(u)v\rangle, \] where $\langle\ ,\ \rangle$ denotes the
   dual pairing; \item[(iii)] Defining $F:B_1\times B_2^*\to\bR$,
   $F(u,\lambda)=f(u)-\langle\lambda,K(u)\rangle$, there is $\lambda\in
   B_2^*$ such that $DF(u,\lambda)(v,\mu) = 0$, for all $v\in B_1, \mu\in
   B_2^*$.  \end{itemize}
\end{Theorem}

We can paraphrase $(i)$ by saying that ``$u$ is a critical point of $f$ on 
$K^{-1}(0)$''. The conditions on $DK$ ensure that $K^{-1}(0)$ is a Banach 
submanifold of $B_1$, by the Implicit Function Theorem, and thus 
$T_u(K^{-1}(0)) = \ker DK(u)$. Clearly, $\lambda$ is the infinite 
dimensional Lagrange multiplier. 

\Proof
  The equivalence of $(ii)$ and $(iii)$ is obvious, as is the implication 
  $(ii)\Rightarrow (i)$.  If $u$ is a critical point of $f$ on $K^{-1}(0)$ 
  then $\ker DK(u) \subset \ker Df(u)\subset B_1$, with both subspaces 
  closed and having closed complements.  It follows that there is a 
  natural projection 
  \[
    \pi:B_1/\ker DK(u) \to B_1/\ker Df(u)
  \]
  which is a bounded map of Banach (quotient) spaces.  Since $Df(u)\in 
  B_1^*$, we have a homomorphism $j_1:B_1/\ker Df(u)\to \bR$.  Since 
  $DK(u)$ is surjective and splits, it factors as $DK(u)=j_2\circ\pi_2$, 
  where $\pi_2:B_1\to B_1/\ker DK(u)$ and $j_2:B_1/\ker DK(u)\to B_2$ is 
  an isomorphism.  Then $\lambda=j_1\circ\pi\circ j_2^{-1}:B_2\to \bR$ is 
  a bounded linear map, ie.~$\lambda\in B_2^*$, and $\lambda\circ DK(u) = 
  j_1\circ\pi\circ\pi_2 = Df(u)$, which gives $(ii)$.
\qed

To show $(ii)\Rightarrow (i)$ in Theorem \bref{Ecrit:thm}, notice that 
for $(g,\pi)\in\cC(\varepsilon,S)$, we have $\cH(g,\pi;\xi)= E(g,\pi) 
- \int_\cM (\xi^0\varepsilon + \xi^i S_i)$ and thus 
\[
  D_{(g,\pi)}\cH(g,\pi;\xi)(h,p) = DE(g,\pi)(h,p),
      \quad\forall (h,p)\in T_{(g,\pi)}\cC(\varepsilon,S).
\]
But $(ii)$ and Theorem \ref{HRT:thm} together imply that 
\[
  D_{(g,\pi)}\cH(g,\pi;\xi)(h,p) = 0\quad \forall (h,p)\in \cG\times\cK,
\]
and $(i)$ follows.
To show the converse $(i)\Rightarrow (ii)$, choose any 
$\tilde{\xi}\in\xinf+W^{2,2}_{-1/2}(\cT)$ and consider the functional 
\[ 
  \tilde{H}(g',\pi') := \cH(g',\pi';\tilde{\xi}),\quad (g,\pi)\in\cF.
\]
From $(i)$ it follows that $(g,\pi)$ is a critical point for both 
$\tilde{H}$ and $E = \xinf^\alpha\bPadm_\alpha$ on the submanifold 
$\cC(\varepsilon,S)$.  We may apply Theorem \ref{LagrangeMult:thm} with 
$B_1=\cG\times\cK\supset\cF$, $B_2=\cL^*$, $K=\Phi-(\varepsilon,S)$ and 
$f=\tilde{H}$; since $(i)$ holds, there is 
$\lambda\in\cL=L^{2}_{-1/2}(\cT)$ such that 
\begin{equation}  \label{Dht:eqn}
          D\tilde{H}(g,\pi)(h,p) = \int_\cM \lambda^\alpha D\Phi_\alpha(g,\pi)(h,p)
\end{equation}
for all $(h,p)\in\cG\times\cK=T_{(g,\pi)}\cF$.
Defining $\xi=\tilde{\xi} + \lambda \in \xinf+L^2_{-1/2}(\cT)$ and 
inserting the definition of $\tilde{H}$ into \bref{Dht:eqn} shows that
$D_{(g,\pi)}\cH(g,\pi;\xi) = 0$;  Theorem \ref{HRT:thm} then implies 
$D\Phi(g,\pi)^*\xi=0$ (weakly) and thus (by Proposition \ref{ker-reg:ppn})
it follows that $\xi\in\xinf+W^{2,2}_{-1/2}(\cT)$ is a generalised 
Killing vector, as required.  This completes the proof of Theorem 
\ref{Ecrit:thm}.
\qed

Observe that under the conditions of Theorem \ref{Ecrit:thm},
alternative $(iii)$ of Theorem \ref{LagrangeMult:thm} shows that
$(g,\pi;\xi)$ is a critical point in all $\cF\times\cL$ for the
functional
\[ 
   \cH(g,\pi;\xi) - \int_\cM(\xi^0\varepsilon+\xi^iS_i).
\]


\begin{thebibliography}{10}

\bibitem{ArmsMarsdenMoncrief82}
J.~Arms, J.~E. Marsden, and V.~Moncrief.
\newblock The structure of the space of solutions of {E}instein's equations
  {II}: Several {K}illing fields and the {E}instein-{Y}ang-{M}ills equations.
\newblock {\em Annals Phys.}, 144(1):81--106, November 1982.

\bibitem{ADM61}
R.~Arnowitt, S.~Deser, and C.~Misner.
\newblock Coordinate invariance and energy expressions in general relativity.
\newblock {\em Phys. Rev.}, 122:997--1006, 1961.

\bibitem{ADM62}
R.~Arnowitt, S.~Deser, and C.~Misner.
\newblock The dynamics of general relativity.
\newblock In L.~Witten, editor, {\em Gravitation}, pages 227--265. Wiley, N.Y.,
  1962.

\bibitem{BahouriChemin99}
H.~Bahouri and J.-Y. Chemin.
\newblock \'equations d'ondes quasilin\'eaires et estimations de {S}trichartz.
\newblock {\em Am. J. Math.}, 121:1337--1377, 1999.

\bibitem{Bartnik86}
R.~Bartnik.
\newblock The mass of an asymptotically flat manifold.
\newblock {\em Comm. Pure Appl. Math.}, 39:661--693, 1986.

\bibitem{Bartnik89}
R.~Bartnik.
\newblock New definition of quasilocal mass.
\newblock {\em Phys. Rev. Lett.}, 62(20):2346--2348, May 1989.

\bibitem{BartnikChrusciel04}
R.~Bartnik and P.~Chru\'sciel.
\newblock Boundary value problems for {D}irac-type equations.
\newblock {\em J. reine u. angewandte Mathematik}, 2004.
\newblock to appear. Math.DG/0307278.

\bibitem{Bartnik02}
R.~Bartnik.
\newblock Mass and 3-metrics of non-negative scalar curvature.
\newblock In Li~Tatsien, editor, {\em Proceedings, ICM 2002}, volume~II, pages
  231--240. HEP Beijing, 2002.
\newblock arXiv/math.DG/0304259.

\bibitem{BeigChrusciel96a}
R.~Beig and P.~T. Chru\'sciel.
\newblock Killing vectors in asymptotically flat space--times: I.
  {A}symptotically translational {K}illing vectors and the rigid positive
  energy theorem.
\newblock {\em J. Math. Phys.}, 37:1939--1961, 1996.

\bibitem{BrillDeserFadeev68}
D.~Brill, S.~Deser, and L.~Fadeev.
\newblock Sign of gravitational energy.
\newblock {\em Phys. Lett.}, 26A(11):538--539, April 1968.

\bibitem{ChoquetYork79}
Y.~Choquet-Bruhat and J.~W. York.
\newblock The {C}auchy problem.
\newblock In A.~Held, editor, {\em General Relativity and Gravitation -- the
  {E}instein Centenary}, chapter~4, pages 99--160. Plenum, 1979.

\bibitem{ChristodoulouOMurchadha81}
D.~Christodoulou and N.~\'o Murchadha.
\newblock The boost problem in general relativity.
\newblock {\em Comm. Math. Phys.}, 80:271--300, 1981.

\bibitem{Chrusciel86a}
P.~T. Chru\'sciel.
\newblock Boundary conditions at spatial infinity from a {H}amiltonian point of
  view.
\newblock In P.~G. Bergmann and V.~de~Sabbata, editors, {\em Topological
  properties and global structure of space-time}. Plenum, New York, 1986.

\bibitem{Corvino00}
J.~Corvino.
\newblock Scalar curvature deformation and a gluing construction for the
  {E}instein constraint equations.
\newblock {\em Commun. Math. Phys.}, 214:137--189, 2000.

\bibitem{CorvinoSchoen03}
J.~Corvino and R.~Schoen.
\newblock On the asymptotics for the vacuum {E}instein constraint equations.
\newblock gr-qc/0301071.

\bibitem{Einstein16}
A.~Einstein.
\newblock Die {G}rundlage der allgemeinen {R}elativit\"atstheorie.
\newblock {\em Annalen d. Phys.}, 49:769, 1916.

\bibitem{FischerMarsden79}
A.~E. Fischer and J.~E. Marsden.
\newblock Topics in the dynamics of general relativity.
\newblock In J.~Ehlers, editor, {\em Structure of Isolated Gravitating
  Systems}, pages 322--395. 1979.

\bibitem{GilbargTrudinger77}
D.~Gilbarg and N.~Trudinger.
\newblock {\em Elliptic Partial Differential Equations of Second Order}.
\newblock Springer Verlag, 2 edition, 1977.

\bibitem{Hilbert15}
D.~Hilbert.
\newblock Grundlagen der physik.
\newblock {\em Nachr. Ges. Wiss. G\"otingen}, page 395, 1915.

\bibitem{HillePhillips57}
E.~Hille and R.~Phillips.
\newblock {\em Functional analysis and semi-groups}, volume~31 of {\em
  Colloquium Series}.
\newblock Am. Math. Soc., 2 edition, 1957.

\bibitem{KlainermanRodnianski02}
S.~Klainerman and I.~Rodnianski.
\newblock Rough solutions to the {E}instein vacuum equations.
\newblock {\em Annals of Math. (submitted)}, 2002.
\newblock math.AP/0109173.

\bibitem{Moncrief75}
V.~Moncrief.
\newblock Spacetime symmetries and linearization stability of the {E}instein
  equations {I}.
\newblock {\em J. Math. Phys.}, 16(3):493--498, March 1975.

\bibitem{Moncrief76}
V.~Moncrief.
\newblock Spacetime symmetries and linearization stability of the {E}instein
  equations {II}.
\newblock {\em J. Math. Phys.}, 17(10):1893--1902, October 1976.

\bibitem{OMurchadha86}
N.~\'O Murchadha.
\newblock Total energy momentum in general relativity.
\newblock {\em J. Math. Phys.}, 27:2111--2128, 1986.

\bibitem{Nester83}
J.~M. Nester.
\newblock The gravitational {H}amiltonian.
\newblock In F.~J. Flaherty, editor, {\em Asymptotic behaviour of mass and
  space-time geometry (Oregon 1983)}, Lecture Notes in Physics 212, pages
  155--163. Springer Verlag, 1984.

\bibitem{ReggeTeitelboim74}
T.~Regge and C.~Teitelboim.
\newblock Role of surface integrals in the {H}amiltonian formulation of general
  relativity.
\newblock {\em Ann. Phys.}, 88:286--318, 1974.

\bibitem{SchoenYau79}
R.~Schoen and S.-T. Yau.
\newblock Proof of the positive mass theorem.
\newblock {\em Comm. Math. Phys.}, 65:45--76, 1979.

\bibitem{SchoenYau81}
R.~Schoen and S.-T. Yau.
\newblock Proof of the positive mass theorem {II}.
\newblock {\em Comm. Math. Phys.}, 79:231--260, 1981.

\bibitem{Stein70}
E.~Stein.
\newblock {\em Singular Integrals and Differentiability Properties of
  Functions}.
\newblock Princeton UP, 1970.

\bibitem{Tataru02}
Daniel Tataru.
\newblock Nonlinear wave equations.
\newblock In Li~Tatsien, editor, {\em Proceedings, ICM 2002}, volume III, pages
  209--220. HEP Beijing, 2002.
\newblock arXiv/math.AP/0304397.

\bibitem{Taylor81}
M.~E. Taylor.
\newblock {\em Pseudodifferential Operators}.
\newblock Princeton UP, 1981.

\bibitem{Witten81}
E.~Witten.
\newblock A simple proof of the positive energy theorem.
\newblock {\em Comm. Math. Phys.}, 80:381--402, 1981.

\bibitem{York80}
J.~York.
\newblock Energy and momentum of the gravitational field.
\newblock In F.~Tipler, editor, {\em Essays in general relativity in honour of
  {A.~T}aub}, pages 39--58. Academic Press, 1980.

\end{thebibliography}

\end{document}